\documentclass[nohyper,twoside]{JHEP35}

\usepackage{graphicx,amsmath,amssymb}

\newcommand{\be}{\begin{equation}}
\newcommand{\ee}{\end{equation}}
\newcommand{\ba}{\begin{eqnarray}}
\newcommand{\ea}{\end{eqnarray}}
\newcommand{\bdm}{\begin{displaymath}}
\newcommand{\edm}{\end{displaymath}}
\newcommand{\beas}{\begin{eqnarray*}}
\newcommand{\eeas}{\end{eqnarray*}}

\newcommand{\av}[1]{\left< #1\right>}

\newcommand{\Veff}{V_{\rm eff}}

\newcommand{\m}{m_\phi}
\newcommand{\M}{M_{\rm Pl}}
\newcommand{\D}{\Delta\phi}

\title{Do Bound Structures Brake Cosmic Acceleration?}
\author{Lily Schrempp and Iain Brown\\
  Institut f\"ur Theoretische Physik, Philosophenweg 16, 69120, Germany\\
 Email: \email{L.Schrempp@thphys.uni-heidelberg.de}, \email{I.Brown@thphys.uni-heidelberg.de}}

\received{\today}

\abstract{In this paper we investigate the impact of the coupling between a quintessence field and clustered matter on the average equation of state of the scalar field. We take the NFW profile to be characteristic of bound structures on galactic and cluster scales, and the isothermal distribution to hold for objects on supercluster scales. Solving analytically for the scalar-field profile, we find that the greatest impact on the equation of state comes from the superclusters. Employing numerical case studies, we verify this effect and probe its dependence on the evolutionary state of the supercluster. An estimate across the Hubble volume yields corrections to the homogeneous equation of state of $\sim$3\%, increasing with coupling strength.}

\preprint{HD-THEP-09-29}

\begin{document}

\section{Introduction}
The discovery of cosmic acceleration requires a source of new physics which may come as a modification to gravity, but is entirely consistent with a cosmological constant or quintessence, a dynamical dark energy~\cite{Wetterich:1987fm,Ratra:1987rm,Caldwell:1997ii}.\footnote{Other suggested alternatives include the backreaction of inhomogeneities on the global expansion rate \cite{Buchert:2001sa}; recent implementations of this have individually produced a small dark energy \cite{Brown:2009cy} and a large variance in the Hubble rate \cite{Clarkson:2009hr} but not provided a unique dark energy source.} Current observations suggest that dark energy comprises approximately 70\% of the energy content of the universe \cite{Komatsu:2008hk,Dunkley:2008ie}. 

The second major ingredient of modern cosmology is the dark matter, making up approximately 25\% of the universe's mass \cite{Komatsu:2008hk,Dunkley:2008ie}. Yet, remarkably, the existence of both of these dominant components of the universe is only deduced from their gravitational interactions. While a fundamental microscopic picture is still lacking, a unified description of the dark sector might exist.

One way to test such a proposition is to explore the phenomenological consequences of possible further non-gravitational interactions between a dynamical dark energy component, an ultra-light quintessence field, and dark matter~\cite{Damour:1990tw,Friedman:1991dj,Gradwohl:1992ue,Anderson:1997un,Carroll:1998zi,Amendola:2001rc,Farrar:2003uw,Gubser:2004du,Bertolami:2004nh,Nusser:2004qu,Bean:2007ny}. As an immediate consequence of the interaction, the scalar field mediates a long-range force between the dark matter particles. As a result, for some scenarios strong constraints on the strength $\beta$ of this dark force have been derived from dark matter dynamics~\cite{Kesden:2006zb,Kesden:2006vz} and from the modified evolution of cosmological perturbations and subsequent impact on the CMB~\cite{Bean:2008ac}, constraining the coupling to $5\%-20\%$ of gravitational strength.

On the other hand, there are several observational motivations for dark interactions beyond gravity which challenge the $\Lambda$CDM concordance model. These include anomalously large bulk flows on scales of $100 h^{-1}\rm{Mpc}$ and beyond \cite{Ayaita:2009qz}, and higher supercluster densities \cite{Einasto:2006vx} and sparser voids \cite{Nusser:2004qu} than expected.

Certain types of adiabatic coupled dark energy scenarios~\cite{Khoury:2003aq,Khoury:2003rn,Brax:2004qh} have been shown to naturally allow for strong couplings $\beta \gg 1$~\cite{Mota:2006ed,Mota:2006fz,Ahlers:2007st,Brax:2007hi} between a light scalar and \emph{all} forms of matter -- even baryons -- by means of the so-called thin-shell or chameleon mechanism. This applies if the field is kept in a minimum of its effective potential inside an overdense region and implies that deviations from the gravitational force law are shielded in the interior of bound objects. 

Another alternative for scenarios with couplings beyond gravity recently studied in the literature is coupled neutrino quintessence~\cite{Fardon:2003eh,Peccei:2004sz,Fardon:2005wc,Amendola:2007yx,Wetterich:2007kr}. The scalar/neutrino interaction results in growing neutrino masses which can trigger the onset of cosmic acceleration~\cite{Wetterich:2007kr}. 

Since all viable dark energy models must yield extremely similar background behaviour, there has been intense study into alternative methods to discriminate between individual homogeneous and inhomogeneous models at a perturbative level. An uncoupled quintessence field with negligible self-interactions is not sensitive to the presence of inhomogeneities in the universe, and thus remains almost perfectly homogeneous. However, within overdensities the presence of a scalar/matter coupling causes the field to collapse along with the dark matter which itself feels an additional fifth force. The resulting modified growth of perturbations has been analysed in the linear~\cite{Afshordi:2005ym,Bjaelde:2007ki,Pettorino:2009vn,Wetterich:2009qf} and in the non-linear~\cite{Nunes:2004wn,vandeBruck:2005ii,Maor:2005hq,Nunes:2005fn,Aghanim:2008zx,Baldi:2008ay,Mota:2008ne,Abramo:2009ne,Saracco:2009df,Wintergerst:2009fh} regimes.

However, during structure formation, the dark energy provides a subdominant component to the energy budget of the universe, thereby limiting the size of the predicted effects. At late times, not only does the dark energy dominate the universe, but the matter distribution is significantly non-linear. It then seems natural to expect further sizeable effects at a non-perturbative level, emerging from non-standard scalar/matter interactions in the recent universe. In this paper, we therefore focus on possible corrections to the homogeneous equation of state of a coupled scalar field resulting from its non-trivial spatial profile acquired within clumped matter.

As a result of its formation and collapse, a gravitationally bound structure in the universe is characterised by a non-trivial density profile with a shape which reflects its evolutionary state. By the current epoch, galaxies and galaxy clusters have had time to develop into fully virialised structures and, according to both observations and simulations, are well-described by a Navarro-Frenk-White density profile. Even though these objects have a sizeable abundance, their total volume makes up a relatively small fraction of the Hubble volume. While it is certainly conceivable that local fluctuations in the coupled quintessence field could produce noticeable corrections to the homogeneous equation of state from such objects, we might expect the greatest corrections to result from the largest and most massive known structures in the universe, the superclusters. However, in contrast to fully relaxed objects on smaller scales, superclusters are still currently in the process of forming. This implies that the spherically averaged density profile of a supercluster depends on its evolutionary state. In the future, it will tend for all superclusters towards an isothermal form proportional to $r^{-2}$ at all radii~\cite{ArayaMelo:2008wk}. However, at present the matter profiles of less evolved superclusters are typically shallower at small radii, but exhibit a steeper slope at large radii~\cite{Wray:2006tw,ArayaMelo:2008wk}.

Our first aim in this paper is to analytically solve the scalar field equations for different kinds of field potential and sourced by matter comprising of isolated, compact objects with density profiles of Navarro-Frenk-White and of isothermal form, corresponding to galaxies and clusters, and to superclusters respectively. The resulting spatial profiles of the field will allow us to determine the average equation of state of the scalar field within such bound structures.

Our second aim is to carefully consider the impact of the evolutionary state of superclusters on the average equation of state of a coupled quintessence field. For this purpose, we will employ the results of numerical simulations to consider two case studies in different stages of formation. We will compare our numerical findings with those for an analytical fit to the simulated data, namely a Gaussian profile in the core stitched to an extended isothermal tail.   

Taken across the Hubble volume, these results will allow us to estimate the overall effect on the global equation of state resulting from the induced inhomogeneous nature of the coupled scalar field.

After setting the stage for our analysis in \S\ref{sec:settingthestage}, we define the average equation of state of an inhomogeneous scalar field. In \S\ref{sec:Dynamics} we provide a model-independent qualitative discussion of the scalar field dynamics in the presence of a coupling to clustered matter. In \S\ref{NFW} and \S\ref{IS} we analytically determine the spatial profile of a scalar field coupled to bound matter structures with an NFW and an isothermal one profile, respectively, and calculate its average equation of state within such objects. In \S\ref{SCprofile} we construct an analytical model for simulated density profiles of representative superclusters in different evolutionary states. In \S\ref{SCcases} we will discuss and compare our analytical and numerical results and estimate the implications on the equation of state averaged across the Hubble volume. \S\ref{Outlook} contains an application of our results to coupled quintessence scenarios with interactions beyond gravity and in \S\ref{Conclusions} we conclude.

\section{Setting the Stage}
\label{sec:settingthestage}
Our aim is to investigate the dependence of the average equation of state of a quintessence field $\phi$ on its coupling to clustered matter.

\subsection{The Matter Distribution}
\label{sec:matterdist}
In the literature, various viable dark energy scenarios have been considered in which a light scalar field couples either to dark matter~\cite{Damour:1990tw,Friedman:1991dj,Gradwohl:1992ue,Anderson:1997un,Carroll:1998zi,Amendola:2001rc,Farrar:2003uw,Gubser:2004du,Bertolami:2004nh,Nusser:2004qu,Bean:2007ny}, to neutrinos~\cite{Fardon:2003eh,Peccei:2004sz,Fardon:2005wc,Amendola:2007yx,Wetterich:2007kr} or democratically to all forms of matter~\cite{Khoury:2003aq,Khoury:2003rn,Brax:2004qh}. In general, all of these types of coupled matter have by the present time developed into bound compact objects and were shown to exhibit similar characteristic density profiles~\cite{Baldi:2008ay,Baldi:2009yv,Wintergerst:2009fh}. In order to deduce results valid for different kinds of coupled quintessence scenarios, we will therefore model the matter distribution as individual, isolated, spherically-symmetric clumps of a variety of sizes embedded in a background vacuum; we will consider bound objects of galactic size, as well as clusters of galaxies and superclusters. This is in contrast to the usual approach where overdensities are embedded in a constant background density. This implies that we will neglect non-linear interactions between the objects; since the typical distance between each of these bound objects is typically significantly larger than their size, this can be justified as a first approximation.

We are likewise taking spherical symmetry to hold for objects across a wide range of scales; for virialised objects up to cluster scales we will assume a Navarro-Frenk-White (NFW) profile~\cite{Navarro:1996gj}, while for superclusters we investigate the characteristic isothermal profile along with modifications to it at small and large radii (see~\cite{Wray:2006tw,ArayaMelo:2008wk} and references therein). Spherical symmetry is a good approximation on smaller scales where objects are fully virialised. When averaging across a large number of objects, we can assume it can also be justified on larger scales, and we would expect our calculations to provide reliable order-of-magnitude estimates.

For the purposes of this paper it suffices to approximate the geometry of space as Minkowski, $g_{\mu \nu}=\eta_{\mu\nu}={\rm diag}(-1,1,1,1)$, implying that both the Newtonian potential and the backreaction due to the energy density in $\phi$ stay small everywhere. Even though we will be taking averages across the Hubble volume, this approximation is valid since we take the bound objects to be isolated from one-another; each resulting average is therefore taken only across a patch of space small compared to the Hubble volume.

In addition, we assume the field dynamics to be static, since the characteristic time scale for a dark energy candidate is of the order of the Hubble time. Furthermore, in order to probe the dependence of the local profile of $\phi$ on the fundamental potential of the scalar field, we will consider two fiducial examples, an inverse-power law and an exponential potential.

\subsection{The Average Equation of State}
\label{sec:averagew}
In this section, we isolate the relevant quantities that define the average equation of state of a scalar field $\phi$ with a self-interaction potential $V(\phi)$. Since an interaction with clustered matter will induce local variations in the field value, we do not {\it a priori} assume that spatial gradients vanish under averaging as they do in the usual homogeneous approximation.

When considering an ``average equation of state'', there are three definitions we can readily imagine. The equation of state of a scalar field is defined from its pressure and energy density, which themselves depend directly on the scalar field itself. We can therefore construct an average equation of state $\overline{w}_1$ by forming the local equation of state and averaging this across a spatial volume; or we can average the scalar field $\phi$ itself and define the average equation of state $\overline{w}_2$ from the general forms of the pressure and energy density; or we can average the pressure and energy density independently and define an average equation of state $\overline{w}_3$ from these. The three definitions can be written
\be
  \overline{w}_1=\overline{\left(\frac{p_\phi(\phi)}{\varrho_\phi(\phi)}\right)}\, , \qquad
  \overline{w}_2=\frac{p_\phi\left(\overline{\phi}\right)}{\varrho_\phi\left(\overline{\phi}\right)}\, , \qquad
  \overline{w}_3=\frac{\overline{p_\phi(\phi)}}{\overline{\varrho_\phi(\phi)}}
\ee
and they coincide for a homogeneous field. Since the relevant thermodynamic quantities are the pressure and the energy density rather than the scalar field or the local equation of state, we choose the last approach.

Varying the scalar field Lagrangian with respect to the metric and averaging over the Hubble volume $\mathcal{V}_H$, the average pressure $\overline{p}_{\phi}$ and energy density $\overline{\varrho}_{\phi}$ of the field can be defined as
\ba
  \overline{p}_{\phi}&=&\mathcal{V}^{-1}_H\int {\rm d}^3 \mathbf{x}\left[\frac{1}{2}\dot{\phi}^2-\frac{1}{6}\left(\nabla\phi\right)^2-V(\phi)\right]
    =\frac{1}{2}\overline{\dot{\phi}^2}-\frac{1}{6}\overline{\left(\nabla\phi\right)^2}-\overline{V(\phi)}\, ,\nonumber\\
  \overline{\varrho}_{\phi}&=&\mathcal{V}^{-1}_H\int {\rm d}^3 \mathbf{x}\left[ \frac{1}{2}\dot{\phi}^2+\frac{1}{2}\left(\nabla\phi\right)^2+V(\phi)\right]
    =\frac{1}{2}\overline{\dot{\phi}^2}+\frac{1}{2}\overline{\left(\nabla\phi\right)^2}+ \overline{V(\phi)}\label{eq:rhopav}\, .
\ea
In the static limit $\dot{\phi}\simeq 0$, the average equation of state $\overline{w}$ therefore takes the form
\be
  \overline{w}=\frac{\overline{p}_{\phi}}{\overline{\varrho}_{\phi}}\simeq \frac{-1-\frac{1}{6}\overline{Q}}{1+\frac{1}{2}\overline{Q}}\label{eq:wav}\, ,
\ee
where
\be
  \overline{Q}\simeq\frac{\overline{\left(\nabla\phi\right)^2}}{\overline{V(\phi)}}\, .
\ee
Accordingly, we can already isolate two exact limiting cases: for $\overline{Q}\gg 1$ the scalar field dynamics will be gradient dominated, leading to $\overline{w}\rightarrow -1/3$; conversely, for $\overline{Q}\ll 1$ the field gradient is subdominant compared to the potential, and $\overline{w}\rightarrow -1$. This implies, therefore, that in the static limit $-1\leq \overline{w} \leq -1/3$.

Gradients of the scalar field originating from an interaction with clustered matter can thus, in principle, lead to sizeable corrections to $\overline{w}$ as compared to its uncoupled, homogeneous analogue $\overline{w}_h$,
\be
  \overline{w}_h=\frac{p_{\phi}\left(\overline{\phi}\right)}{\varrho_{\phi}\left(\overline{\phi}\right)}
  \simeq\frac{\frac{1}{2}\left(\dot{\overline{\phi}}\right)^2-V\left(\overline{\phi}\right)}{\frac{1}{2}\left(\dot{\overline{\phi}}\right)^2+V\left(\overline{\phi}\right)}
  \simeq -1\label{eq:wavh}\, ,
\ee
where again the last equality applies in the static limit.

To transfer the calculations for the equation of state of the coupled field within a bound object to an average across the Hubble volume it is necessary to make some assumptions concerning the distribution of matter sources. We will separate the bound objects into distinct classes, each of which produces a particular scalar field profile and therefore the same equation of state. If $i$ denotes a particular class of objects with radius $r_{{\rm obj}, i}$, then Eq.~(\ref{eq:wav}) can be approximated by
\be
  \overline{w}\simeq=\frac{-1-\frac{1}{6}\sum\limits ^{}_iQ_i\mathcal{R}_i}{1+\frac{1}{2}\sum\limits ^{}_iQ_i\mathcal{R}_i}\, ,
\ee
with
\be
  \mathcal{R}_i=N_i\frac{\mathcal{V}_{{\rm obj},i}}{\mathcal{V}_{H}}\, , \qquad
  Q_i=\frac{\av{(\nabla\phi)^2}_i}{\av{V(\phi)}_i}
       =\frac{\int\limits^{r_{{\rm obj},i}}_{0}{\rm d}rr^2\left(\nabla\phi(r)\right)^2}{\int\limits^{r_{{\rm obj},i}}_{0}{\rm d}rr^2V(\phi(r))}\label{eq:wavsum}\, .
\ee
Here $N_i$ is the number of objects in the class $i$, we have assumed $Q\simeq 0$ in the intergalactic medium, and angle brackets refer to quantities averaged over the volume of an isolated object $\mathcal{V}_{{\rm obj},i}$. Note that $\mathcal{R}_i$ is a measure for the volume dilution of the contribution $Q_i$ to the sum. We can then immediately deduce that notable differences between Eq.~(\ref{eq:wavsum}) and Eq.~(\ref{eq:wavh}) could well emerge from a class of objects for which the gradient contribution $Q_i$ is large enough to compensate for the volume dilution implied in $\mathcal{R}_i$.

\section{Dynamics of the Scalar Field}
\subsection{The Equation of Motion}
\label{sec:Dynamics}
As is clear from the previous section, to estimate $\overline{w}$ we must determine the spatial profile of the scalar field $\phi$ resulting from its interaction with clustered matter. While we assume standard dark matter density distributions for the coupled matter species, we will consider different possible forms of coupling functions leading to distinct dynamics of the field. By this means our results can be generalised to hold for the various different kinds of coupled scalar field scenarios mentioned in \S\ref{sec:settingthestage}. 

In general, in the presence of an interaction of strength $\beta_a$ between $\phi$ and $a\geq 1$ non-relativistic matter species, the dynamics of $\phi$ are governed by an effective potential of the form
\be
  V_{\rm eff}(\phi,r)=V(\phi,r)+\sum \limits^{}_a \rho_a(\phi,r)\, ,
\ee
where the first term arises from self-interactions of the quintessence field, generating a monotonically-decreasing potential $V(\phi)$ of a runaway form. The second contribution to $V_{\rm eff}$ originates from the $\phi$ dependence of the masses of the coupled species, $m_a(\phi)$, introduced by the interaction. Since the corresponding energy density $\rho_a$ is a function of the particle mass, as a direct consequence $\rho_a(m_a(\phi))$ contributes to the effective potential for $\phi$.

For simplicity, we assume a universal coupling of strength $\beta=\beta_a$ and an effective energy density $\rho$ leading to an effective potential of the form
\be
  V_{\rm eff}(\phi,r)=V(\phi,r)+\rho(\phi,r)\label{eq:Veff},
\ee
which encodes the effects of one or more coupled species on the scalar field potential.

Note that depending on the sign of the coupling function ${\rm d}\log(m(\phi))/{\rm d}\phi$, we must in general discriminate between two cases which lead to distinct dynamics. Namely, a positive sign allows the effective potential to exhibit a minimum, while a negative sign implies a (monotonic) decrease of $V_{\rm eff}$ with $\phi$.

We take for simplicity the mass of the coupled matter to depend only very weakly on changes in the scalar field value such that it can be taken to be approximately constant even within environments with strongly varying matter density -- that is, $\rho(\phi,r)\simeq \rho(r)$. Under the further assumption of spherical symmetry for the matter profile, the dynamics of the field in the static limit are described by the equation of motion
\be
  \phi^{\prime \prime}(r)+\frac{2}{r}\phi^{\prime}(r)=V_{,\phi}(\phi,r)+\beta\rho(r)=V_{\rm{eff},\phi}(\phi,r)\, ,
  \label{eq:poisson}
\ee
where here and in the following primes denote derivatives with respect to $r$, a subscript ${,\phi}$ is a derivative with respect to $\phi$, and the last equality results from Eq.~(\ref{eq:Veff}) and a coupling function of exponential form.

It will transpire that our results are not sensitive to the form of the potential $V(\phi)$. We demonstrate this by employing two fiducial examples. The first is of exponential form,
\be
  V(\phi)=\M^{4}e^{-\frac{\alpha}{\M}\phi} \label{eq:Vexp}\, ,
\ee
where $\M$ denotes the reduced Planck mass and $\alpha\gtrsim 10$ to fit observation~\cite{Doran:2007ep}. 

Our second fiducial potential is an inverse power law potential of the form
\be
  V(\phi)=\frac{M^{4+n}}{\phi^n} \label{eq:Vinv}\, ,
\ee
where $n$ is a positive constant and $M$ an intrinsic mass scale. It fixes the behaviour of $\phi$ in the intergalactic medium and has to be compared to the scale of dark energy in a homogeneous universe. 

However, unless specified the potential should be taken to be general.

\subsection{Qualitative Description of the Solutions}
\label{Quali}
Let us start our analysis with a model-independent description of different possible kinds of approximate analytical solutions to Eq.~(\ref{eq:poisson}).

We first describe the initial conditions that we assume to hold for any effective potential. Since the field equation in Eq.~(\ref{eq:poisson}) is a second order differential equation, we have to specify two boundary conditions for $\phi$. Firstly, to guarantee a non-singular behaviour of $\phi$ at the origin, we require as usual
\be
  \phi^{\prime}=0\mbox{ at } r=0\, . \label{eq:ini1} 
\ee
To set the second boundary condition, we note that in our setting the bound matter clumps are embedded in a vacuum, and so the matter density outside of these asymptotes to zero and not to a non-zero background value as it would for an idealised homogeneous universe. We must accordingly take the field to be essentially free in the intergalactic medium, that is $\phi=\phi_{\rm free}$. Therefore, in `empty space' its dynamics are solely governed by its self-interaction potential $V(\phi_{\rm free})$. Our second boundary condition is then
\be
  \phi\rightarrow\phi_{\rm free}\mbox{ as } r\rightarrow \infty\, .
\ee

To a good approximation, the solutions for $\phi$ turn out to be independent of the scalar field potential $V(\phi)$. However, there is one situation in which this is not the case, where $\phi=\phi_{\rm min}$, with $\phi_{\rm min}$ being the field value at the minimum. This can occur when the following conditions are satisfied:
\ba
  &&V_{\rm{eff},\phi}=0,\mbox{ such that $V_{\rm eff}$ exhibits a minimum}\label{eq:Veffmin} \\
  &&\phi_i-\phi_{\rm min}(0)\ll\phi_{\rm min}(0),\mbox{ where $\phi_i$ is the value of $\phi$ at the origin}\label{eq:phi_i} \\
  &&\left|\frac{\rho'}{\rho}\right|\ll m_\phi(r),\mbox{ which we refer to as the adiabaticity condition.}\label{eq:condmin}
\ea
If these conditions hold, then the equations of motions allow for an adiabatic solution to Eq.~(\ref{eq:poisson}), valid in the range $0\leq r\leq r_{\rm kick}$ for $r_\mathrm{kick}\leq r_{\rm obj}$, where the field instantaneously adjusts to spatial changes of the minimum $V_{\rm{eff},\phi}$. Here, $r_{\rm kick}$ denotes the radial coordinate at which the fields gets kicked out of the minimum. The adiabaticity condition corresponds to the requirement that the matter density changes on scales much larger than the Compton wavelength $\m^{-1}$ of the scalar field in that region.

Since in this case the source term is initially $V_{\rm{eff},\phi}\simeq 0$, and the mass of the field inside the object is large, $\m(0)\gg m_{\rm free}$, perturbations in the field will be exponentially suppressed. As a result, if $r\leq r_{\rm kick}$ and Eq.~(\ref{eq:condmin}) is valid, the matter source term $\rho(r)$ can be treated as a constant in the scalar field equation of motion~\cite{Khoury:2003rn}.

As soon as $r>r_{\rm kick}$, condition (\ref{eq:condmin}) is violated and the field gets displaced from the minimum, and the contribution from $V_{,\phi}$ to the field equation, Eq.~(\ref{eq:poisson}), becomes negligible with respect to the matter density $\rho$,
\be
  \frac{\beta \rho(r)}{\M}\gg V_{,\phi}\, . \label{eq:pseudo} 
\ee

Note that if the condition (\ref{eq:phi_i}) is violated, this implies that the field is already sufficiently displaced from the minimum at the origin such that $r_{\rm kick}=0$ and its evolution is non-adiabatic from the start.

As it will turn out, for the density profiles and potentials discussed in this paper, Eq.~(\ref{eq:condmin}) is already violated for very small $r$. Accordingly, the field is immediately kicked out of the minimum of $V_{\rm eff}$ even in scenarios that in principle might obey all three conditions. As a result, the evolution is effectively non-adiabatic across virtually the entire object. Our result agrees with previous numerical studies, which found that the so-called chameleon effect~\cite{Khoury:2003aq,Khoury:2003rn,Brax:2004qh} is not significant for compact structures at low redshifts~\cite{Li:2009sy,Zhao:2009ke,Li:2010mq}. 

In general, therefore, in the scenarios we will consider we are lead to the second type of approximate solution to Eq.~(\ref{eq:poisson}). This solution is valid as long as Eq.~(\ref{eq:pseudo}) is fulfilled and implies that Eq.~(\ref{eq:poisson}) reduces to
\be
  \phi^{\prime\prime}+\frac{2}{r}\phi^{\prime}\simeq \frac{\beta\rho(r)}{\M}\, .\label{eq:rhodom}
\ee
Following the approach described in~\cite{Mota:2006ed}, we have verified that this approximation stays valid for $0\leq r\leq r_{\rm obj}$ in the scenarios under investigation in this paper.

Note for later reference that Eqs.~(\ref{eq:ini1}) and (\ref{eq:rhodom}) yield the following result for the gradient squared of $\phi$,
\be
  (\nabla\phi(r))^2=\left(\frac{\beta}{\M}\right)^2r^{-4}\left(\int\limits_{0}^{r}{\rm d} \tilde{r}\tilde{r}^2\rho(\tilde{r})\right)^2\, ,\label{eq:gradnoint}
\ee
independent of the form of the scalar potential $V(\phi)$.
 
Averaged over the volume $\mathcal{V}_{\rm obj}=4\pi r_{\rm obj}^3/3$ of the object, this gives
\be
  \langle (\nabla \phi)^2 \rangle = 4\pi \mathcal{V}^{-1}_{\rm obj}\left(\frac{\beta}{\M}\right)^2
    \int\limits_{0}^{r_{\rm obj}}{\rm d} rr^{-2}\left(\int\limits_{0}^{r}{\rm d}\tilde{r}\tilde{r}^2\rho(\tilde{r})\right)^2\, .\label{eq:gradgen}
\ee

Finally, outside of the object the dynamics of the field are only governed by its self-interaction potential $V$ with $V_{,\phi}\ll 1$. Accordingly, in this regime the field equation reduces to
\be
  \phi^{\prime\prime}+\frac{2}{r}\phi^{\prime}\simeq 0\, ,
\ee
which gives the solution
\be
  \phi(r)=\phi_{\rm free}-\frac{Ce^{(-m_{\rm free}(r-r_{\rm obj}))}}{r}\, .
\ee
Here we have employed that the Compton wavelength of the field in the intergalactic medium, $m_{\rm free}^{-1}$, is much larger than the radius, $r_{\rm obj}$, of the object. Furthermore, the constant $C$ can be determined by requiring continuity of $\phi$ and ${\rm d}\phi/{\rm d} r$ at $r=r_{\rm obj}$.

\section{NFW matter density profile}
\label{NFW}
We now turn to objects of galactic and cluster scales, for which we employ an NFW matter density profile~\cite{Navarro:1996gj}, which is a phenomenological fit to the density distribution of dark matter halos in CDM cosmological simulations,
\be
  \rho(R)=\frac{\rho_s}{R(1+R)^2}\, ,\label{eq:rhoNFW}
\ee
with
\be
  \rho_s = \frac{\Delta}{3}\frac{c^3\rho_c}{\log(1+c)-c/(1+c)}\, , \qquad c=\frac{r_{\rm vir}}{r_s}\, ,
\ee
and we have defined the dimensionless variable $R\equiv r/r_s$, where $0\leq R \leq c$, with $r$ being the radial coordinate and $r_s$ the characteristic scale of the halo. Together with its virial radius $r_{\rm vir}$, $r_s$ defines the concentration parameter $c$. Furthermore, the characteristic density $\rho_s$ is proportional to the the critical density of the universe, $\rho_c$, and to the virial overdensity, $\Delta$, which can be computed as described in \cite{Bryan:1997dn}. While the two parameters $\rho_s$ and $r_s$ are in principle independent, they where shown in simulations \cite{Wechsler:2001cs} to be correlated through the concentration parameter $c$, the critical density $\rho_c$ and the virial mass ${\rm M}_{\rm vir}$,
\be
  M_{\rm vir}=4\pi\rho_s r_s^3\left[\log(1+c)-\frac{c}{1+c}\right].\label{eq:massNFW}
\ee 
Note that typical virial masses of galaxies and galaxy clusters are $M_{\rm vir}\simeq 10^{10}- 10^{13} M_{\odot}$ and $M_{\rm vir} \simeq 10^{14}- 10^{15} M_{\odot}$, respectively.

In the following, we will employ the relations \cite{Gentile:2006hv}
\be
  c\simeq 13.6\left(\frac{{\rm M}_{\rm vir}}{10^{11}{\rm M}_{\odot}}\right)^{-0.13}\, ,\qquad
  r_s \simeq 8.8\left(\frac{{\rm M}_{\rm vir}}{10^{11}{\rm M}_{\odot}}\right)^{0.46}{\rm kpc}\, .\label{eq:crs}
\ee

As discussed in the last section, the behaviour of the coupled scalar field inside a matter lump crucially depends on the magnitude of its interaction range $\m^{-1}$ compared to the rate of change of the matter density, see Eq.~(\ref{eq:condmin}). 

In the following subsection we derive the range of validity of the adiabatic approximation to the field equation for the two fiducial potentials in Eq.~(\ref{eq:Vexp}) and Eq.~(\ref{eq:Vinv}). In addition, we assume the matter coupling function to be an increasing function of the field value such that the effective potential exhibits a minimum.

\subsection{Adiabaticity Condition for an NFW Matter Profile}
\label{adNFW}
For the NFW distribution the adiabaticity condition (\ref{eq:condmin}) becomes
\be
  \left|\frac{\rho^{\prime}(R)}{\rho(R)}\right|=\frac{1}{r_s}\frac{1+3R}{R(1+R)}\simeq \frac{1}{rs}R^{-1}\, , \label{eq:changeNFW}
\ee
where the last equality assumes $R \ll 1$. As a result of the cusped structure of the matter density, Eq.~(\ref{eq:changeNFW}) diverges at the origin. This already suggests that the field will be immediately displaced from the minimum, as we now explicitly verify.

In the case of the exponential potential, we find small excitations about the minimum of the potential to be given by
\be
  m_{\phi}(R)=\frac{(\alpha\beta\rho_s)^{\frac{1}{2}}}{\M}(1+R)^{-1}R^{-\frac{1}{2}}\, .\label{eq:mexpNFW}
\ee
The resulting adiabaticity condition therefore takes the form
\be
  R\ll -\frac{1}{3}+\frac{1}{18}\left(\frac{\alpha\D_{\rm NFW}}{\M}\right)\left[1+\left(1-\frac{12\M}{\alpha\D_{\rm NFW}}\right)^{\frac{1}{2}}\right]\label{eq:condexpNFW},
\ee
with
\be
  \D_{\rm NFW}\equiv \frac{\beta\rho_s r_s^2}{\M}\simeq \mathcal{O}(10^{-4})\M\left(\frac{M_{\rm vir}}{10^{15}M_{\odot}}\right)^{0.54}\, ,\label{eq:deltaphiNFW}
\ee
where the last equality results from Eqs.~(\ref{eq:massNFW})--(\ref{eq:crs}) and holds for $c\gtrsim 1$. As it will turn out later on, $\D_{\rm NFW}$ is a measure for the absolute change of the scalar field value within the object, which for typical model parameters is much smaller than the characteristic mass scale in the potential, $\D_{\rm NFW}  \ll \M/\alpha$. Eq.~(\ref{eq:condexpNFW}) therefore cannot be solved for any $R\geq 0$ and the field evolution is non-adiabatic from the start.

In the case of the inverse-power law potential, the field mass expanded around the origin takes the form
\be
  m_{\phi}(R) \simeq \sqrt{n+1}n^{-\frac{1}{2}\left(\frac{1}{n+1}\right)}M^{-\frac{1}{2}\left(\frac{n+4}{n+1}\right)}
    \left(\frac{\D_{\rm NFW}}{r_s^2}\right)^{\frac{1}{2}\left(\frac{n+2}{n+1}\right)}R^{-\frac{1}{2}\left(\frac{n+2}{n+1}\right)}\label{eq:minvNFW}\, .
\ee 
Again, for any positive $n$, Eq.~(\ref{eq:minvNFW}) diverges less fast than Eq.~(\ref{eq:changeNFW}) at small radii. 

In this case, the adiabatic condition in Eq.~(\ref{eq:condmin}) close to the origin becomes
\be
  R \ll (n+1)^{\left(\frac{n+1}{n}\right)} n^{-\frac{1}{n}}M^{\frac{n+4}{n}} r_s^{-\frac{2}{n}}(\D_{\rm NFW})^{-\left(\frac{n+2}{n}\right)}\ll 1\, ,
\ee   
where the last equality holds even for the largest and most massive clusters of galaxies and realistic model parameters.

We therefore conclude that Eq.~(\ref{eq:pseudo}) can be assumed to hold and the field evolution is non-adiabatic within objects up to cluster scales at the current epoch. This result is in agreement with corresponding numerical studies that have demonstrated the absence of the thin-shell effect for compact structures at late times. In a series a papers~\cite{Li:2009sy,Zhao:2009ke,Li:2010mq} the authors employed full numerical simulations to analyse the development as well as the disappearance of thin shells in chameleon-like scalar field scenarios as a function of redshift during nonlinear structure formation.

Note also that the resulting solutions for the field thereby become independent of the functional form of the scalar field potential. 

\subsection{Non-Adiabatic Solution for the NFW Matter Profile}
\label{nonadNFW}
For an NFW profile, we find that the solutions to the field equation in Eq.~(\ref{eq:rhodom}) are
\ba
  \phi_{\rm{in}}&=&\phi_{\rm{free}}-\D_{\rm NFW}\left[\frac{\log(1+R)}{R}-\frac{1}{1+c}\right],\label{eq:NFWphi}\label{eq:phiinNFW}\\
  \phi_{\rm{out}}&=&\phi_{\rm{free}}-\D_{\rm NFW}\left(\log(1+c)-\frac{c}{1+c}\right)\frac{e^{-m_{\rm{free}}r_s(R-c)}}{R}\label{eq:NFWphiout}\, ,
\ea
where the result again assumes $m_{\rm free}r_{\rm vir}\ll 1$. Furthermore, as described in \S \ref{Quali}, we have determined the integration constants by matching the inner solution $\phi_{\rm in}$ to the outer solution $\phi_{\rm out}$, at $R=c$ and required continuity of the first derivative of the field at $R=c$.

As we mentioned in the last section, the absolute change in the scalar field value, $\phi(c)-\phi(0) \simeq \D_{\rm NFW}$ defined in Eq.~(\ref{eq:deltaphiNFW}) within a compact object with virial radius $r_{\rm vir}$, is a measure for the total mass $M_{\rm vir}$ defined in Eq.~(\ref{eq:massNFW}), which is interacting with $\phi$ with coupling strength $\beta$. 

Using Eq.~(\ref{eq:phiinNFW}), the volume average of the exponential potential becomes
\be
  \langle V \rangle \simeq V_{\rm free}\left[1+\frac{\alpha\Delta\phi_{\rm NFW}}{\M}\left(\frac{1}{4}+(1-c)\left(2-3\log 2\right)\right)\right]
\ee
for $c\gtrsim 1$.

In the case of the inverse power law potential we find
\be
  \langle V \rangle \simeq V_{\rm free}\left [ 1+n\frac{\D_{\rm NFW}}{\phi_{\rm free}}  \left(\frac{1}{4}+(1-c)\left(2-3\log 2\right)\right)\right]\label{eq:VinvavNFW}
\ee
for $c\gtrsim 1$. Note that for both potentials, the matter coupling results in a positive correction term with respect to the free field potential $V_{\rm free}$. However, importantly, for typical model parameters applying to galaxies and galaxy clusters, it turns out to be negligible, since according to Eq.~(\ref{eq:deltaphiNFW}) $\D_{\rm NFW} \ll \phi_{\rm free}$ and $\D_{\rm NFW}  \ll \M/\alpha$, respectively. 

For the volume average of the gradient squared we get
\ba
  \langle (\nabla\phi)^2 \rangle &=& 3c^{-3}\left(\frac{\D_{\rm NFW}}{r_s}\right)^{2} \int\limits_{0}^{c}{\rm d} R\left( \frac{\log(1+R)}{R}-\frac{1}{1+R}\right)^2\nonumber\\
  &=&3f(c)\left(\frac{\D_{\rm NFW}}{cr_s}\right)^{2}\label{eq:gradNFW}
\ea
with
\be
  f(c)=\left[\frac{1}{1+c}-\left(\frac{\log(1+c)}{c}\right)^2\right],
\ee
where $f(c)\sim \mathcal{O}(10^{-2})$ for typical values of the concentration parameter $c$. Note that even though the matter density formally diverges at the origin, the gradient squared stays well-defined at $R=0$. This is due to fact that the volume integral over the matter density is finite as can be seen from Eq.~(\ref{eq:gradgen}).

Accordingly, using Eq.~(\ref{eq:crs}), the volume average of the gradient squared normalised to the free field potential becomes
\be
  Q=\frac{\langle (\nabla \phi)^2 \rangle}{V_{\rm free}}\simeq 5.12\times 10^{-6} \beta^2 \left(\frac{M_{\rm vir}}{10^{11} M_{\odot}}\right)^{0.68}
\ee
for typical galaxies, and
\be
  Q=\frac{\langle (\nabla \phi)^2 \rangle}{V_{\rm free}}\simeq 1.50\times 10^{-2}\beta^2 \left(\frac{M_{\rm vir}}{10^{15} M_{\odot}}\right)^{0.68}\label{eq:QNFW}
\ee
for higher mass clusters, such that $Q\ll 1$ for couplings of (sub-)gravitational strength $\beta\lesssim 1$ (see, however, \S\ref{Outlook}, where we analyse the results for coupled quintessence scenarios with interactions of super-gravitational strength).

Recall that, according to Eq.~(\ref{eq:wavsum}), contributions to $\overline{w}$ depend on the product of $Q$ and $\mathcal{R}$. However, for all classes of objects on galactic and cluster scales, the volume suppression amounts to at least $\mathcal{R}\lesssim \mathcal{O}(10^{-12})N \ll 1$ for typical abundances of these object classes. $\overline{w}$ therefore receives negligible corrections compared to the uncoupled case $\overline{w}_h\simeq -1$.

As a concluding remark, let us note that the small result for $Q$ is a direct consequence of the specific functional form of the NFW profile. Namely, according to Eq.~(\ref{eq:gradNFW}), the integrand of the volume average is of $\mathcal{O}(R^2)$ for $R\ll 1$ and of $\mathcal{O}(1/R^2)$ for $R\gg 1$. This implies that in both regimes the corresponding contributions to the volume average are strongly suppressed and thus the average gradient squared only receives sizeable contributions from radii $R\sim 1\ll c$.

However, as we will see in the next section, for the density distribution characteristic for objects on supercluster scales, the average field gradient receives almost equal contributions at all radii and thus turns out the be much larger.

\section{Superclusters}
\label{SC}
In this section we investigate the impact of superclusters, the largest known bound structures in the universe, on $\overline{w}$ in coupled quintessence. Within a $\Lambda$CDM cosmology, superclusters are expected to have masses ranging from ${\rm M_{SC}}\sim 10^{15}h^{-1}{\rm M_\odot}$ up to greater than $6\times 10^{16}h^{-1}{\rm M_{\odot}}$, with sizes $\sim 10h^{-1}$--$160 h^{-1}{\rm Mpc}$~\cite{Wray:2006tw,Bolejko:2006vw}. Superclusters therefore fill approximately $3\%-5\%$ of the current Hubble volume~\cite{Wray:2006tw}.

Since structure forms hierarchically, the superclusters we observe in the present universe are still in the process of formation, while galaxies and clusters of galaxies have already developed into collapsed and virialised structures. Accordingly, depending on their evolutionary state, superclusters tend to have a much shallower density profile than fully relaxed objects.   

According to simulations, the spherically averaged density profile of superclusters is well-fitted by an isothermal profile $\rho(r)\propto r^{-2}$ over a broad range in radius. While the profile is shallower than this at low radii, it steepens at larger radii~\cite{Wray:2006tw,ArayaMelo:2008wk}. Since we only have statistical information on the exact density distribution, we will proceed in the following manner.

To establish some intuitive understanding of the behaviour of the field inside a typical supercluster, we will begin by analytically determining the generic profile of a scalar field interacting with isothermally distributed matter. This allows us to get a feeling for the magnitude of the effects on the global equation of state as a function of the coupling, the mass and the radius of a single, standard supercluster.

Subsequently, we will turn to two simulated supercluster case studies. These allow us to test both analytically and numerically the dependence of our results on deviations from the isothermal profile for small and large supercluster radii.

Finally, we will end our discussion of superclusters by estimating the overall effect on the global equation of state.

\subsection{Isothermal Matter Distribution}
\label{IS}
Formally, the isothermal profile $\rho(R)\propto R^{-2}$ exhibits a singularity at the origin, where $R\equiv r/r_{\rm SC}$ denotes the radial coordinate normalised to the supercluster radius $r_{\rm SC}$. However, bound objects which have not yet relaxed will exhibit a flattening of the matter distribution at some $R=A$ with $0<A \ll 1$. Accordingly, for $A\leq R\leq 1$ the isothermal distribution expressed in terms of the finite matter density $\tilde{\rho}\equiv\rho(A)A^2$ reads
\ba
  \rho_{\rm IS}(R)=\frac{\tilde{\rho}}{R^2}\label{eq:rhoIS}\, .
\ea

With the help of Eq.~(\ref{eq:pseudo}), the field equation Eq.~(\ref{eq:poisson}) then takes the form
\be
  \frac{ {\rm d}^2 \phi}{{\rm d} R^2}+\frac{2}{R}\frac{ {\rm d}\phi}{{\rm d} R}\simeq \frac{\beta}{\M}\frac{\tilde{\rho}}{R^2}\, .
\ee
Assuming both the field value and its first derivative to be continuous at $R=A$, we get for $A\leq R\leq 1$ that
\ba
  \phi_{\rm in}&=&\phi_{\rm free}-\D_{\rm IS}\left[1-\log(R)-\frac{A}{R}\left(1-A\D^{-1}_{\rm IS}\left.\frac{{\rm d}\phi}{{\rm d}R}\right|_{R=A}\right)\right],\label{eq:phiSCdiffin}\\
  \phi_{\rm{out}}&=&\phi_{\rm free}-\D_{\rm IS}\left[ 1-A\left(1-A\D^{-1}_{\rm IS}\left.\frac{{\rm d}\phi}{{\rm d}R}\right|_{R=A}\right)\right],\label{eq:phiSCdiffout}
\ea
where
\be
  \D_{\rm IS}=\frac{\beta\tilde{\rho}r_{\rm SC}^2}{\M}.\label{eq:Deltar2}
\ee
As described in \S\ref{sec:Dynamics}, we have determined the integration constants by matching the inner solution $\phi_{\rm in}$ to the outer solution $\phi_{\rm out}$, and by requiring continuity of the respective derivatives at $R=1$.

Note that we can disentangle the different contributions to the average gradient squared by investigating the limiting case $\D_{\rm IS}\gg A\left.{\rm d}\phi/{\rm d}R\right|_{R=A}$. Doing so assumes the profile of $\phi$ in the isothermal region to be insensitive to its slope in the core region; in general, this slope is associated with large uncertainties introduced by the matter profile in the core. Comparing this with the analysis in \S\ref{SCprofile}, where we employ a Gaussian distribution at small radii, will allow us to gain some understanding of the influence of the core profile.

According to Eq.~(\ref{eq:phiSCdiffin}), for $\D_{\rm IS}\gg A\left.{\rm d}\phi/{\rm d}R\right|_{R=A}$ we find the gradient squared of an object with isothermal density profile averaged over the volume of the object to be
\ba
  \langle (\nabla \phi)^2\rangle&=&3\left(\frac{\D_{\rm IS}}{r_{\rm SC}}\right)^2 \int\limits_{A}^{1}{\rm d} R\left(1-\frac{A}{R}\right)^2
    =3\left(\frac{\D_{\rm IS}}{r_{\rm SC}}\right)^2 \left[1+\left(\frac{2\log(A)}{A}-1\right)A^2\right]\nonumber
  \\ &\simeq& 3\left(\frac{\D_{\rm IS}}{r_{\rm SC}}\right)^2\left[1+2\log(A)A+\mathcal{O}(A^2)\right],
\ea
where the last equality assumes $A\ll 1$. Note that in this limit the integrand of the volume average is of order unity for $A\ll R \leq 1$ implying that the average gradient squared receives almost equally large contribution at all radii in this range. Contrasting this result with its analogue in the NFW case in Eq.~(\ref{eq:gradNFW}), we observe that the isothermal matter distribution results in an average gradient squared which is a factor $f^{-1}(c)\sim\mathcal{O}(10^{2})$ larger. We therefore arrive at the crucial result that the impact of the interaction on the coupled scalar field equation of state strongly depends on the radial matter distribution. Since the spherically averaged supercluster profile is typically isothermal over a broad range in radius, this result already suggests that the dominant contribution to the average gradient squared of the field will originate from this region. 

\subsection{Composite Supercluster Profiles}
\label{SCprofile}
In the quantitative investigation that follows, we consider two representative superclusters, SC $8$ and SC $98$, with masses of $M_{\rm SC}\sim 5.4\times 10^{15} h^{-1} M_{\odot}$ and $M_{\rm SC}\sim 3.6\times 10^{15} h^{-1} M_{\odot}$, respectively, taken from a simulation performed on the Beowulf Cluster at the University of Groningen \cite{ArayaMelo:2008wk}. While at the current epoch, SC $8$ is already a centrally concentrated object, SC $98$ is still more extended. 

However, both supercluster density profiles exhibit as a characteristic basic shape a high-density core with a central density of $\sim (10^4-10^5)\rho_{c}$, embedded within an isothermal power law region with slope $-2$.

Furthermore, for both of these examples we find that the central region is well-fitted by a Gaussian density distribution. Since the two examples represent in some respect two extremes of supercluster evolution, it is a reasonable assumption that this is a generic feature. In this section, therefore, we model a generic supercluster density profile by stitching together at low radii a Gaussian and an isothermal matter density profile. This then allows us to determine analytical solutions for the scalar field profile within typical high-mass superclusters which will be compared to our corresponding numerical results for the spherically averaged density profiles of SC $8$ and SC $98$.

\subsubsection{Gaussian Matter Distribution}
\label{Gaussian}
In the following we determine the field profile $\phi_{\rm G}$ arising from a small Gaussian matter region stitched to an extended isothermal tail. The Gaussian and isothermal regions are connected at a small radius $A\ll 1$.

We parameterise the Gaussian density profile by
\be
  \rho_{\rm G}(R)=\rho_{\rm max}e^{-\frac{1}{2}(Rc_g)^2}\, ,\label{eq:rhoG}
\ee
with
\be
  \rho_{\rm max}=\frac{\cal{A}}{(2\pi)^{3/2}}\left(\frac{c_g}{r_{\rm obj}}\right)^3\, .
\ee
We have defined the dimensionless constant $c_g\equiv r_{\rm obj}/\sigma$, $\sigma$ being the half-width of the Gaussian distribution with amplitude $\cal{A}$.  

In the following subsection we derive the range of validity of the adiabatic approximation to the field equation for the two fiducial potentials in Eqs.~(\ref{eq:Vexp}--\ref{eq:Vinv}) and will argue that the results hold for any realistic analytical scalar field potential. 

\subsubsection{Adiabaticity Condition for a Gaussian Core Profile}
\label{adGaussian}
If the effective potential exhibits a minimum and $V(\phi)$ is of exponential form, small perturbations about the minimum are described by
\be
  m_{\phi}(R)=\left(\frac{c_g}{r_{\rm obj}}\right)\sqrt{\frac{\alpha\D_G}{\M}}e^{-\frac{1}{4}(Rc_g)^2}\, .
\ee  
Accordingly, the minimum condition in Eq.~(\ref{eq:condmin}) reads
\be
R \ll \frac{\sqrt{2}}{c_g}e^{-\frac{1}{2}\mathcal{W}\left(\frac{1}{2}\frac{\alpha\D_G}{\M}\right)}\simeq \frac{1}{c_g}\sqrt{\frac{\alpha\D_G}{\M}} \ll 1\, ,\label{eq:minGexp}
\ee 
where the last equalities hold for typical parameter values for which $\alpha\D_G/\M\ll 1$ and $c_g\gg 1$. Here $\mathcal{W}(x)$ is the Lambert function, which satisfies $\mathcal{W}(x)e^{\mathcal{W}(x)}=x$.

If the potential exhibits a minimum and $V(\phi)$ is of inverse power law form, we get
\be
  m_{\phi}(R)=\mathcal{K}\left(\frac{c_g}{r_{\rm obj}}\right)e^{-\frac{1}{4}\left(\frac{n+2}{n+1}\right)(Rc_g)^2}\, ,
\ee
with
\be
  \mathcal{K}=\sqrt{n+1}\left(nM^{n+4}\right)^{-\frac{1}{2(n+1)}}\left(\frac{c_g}{r_{obj}}\right)^{\frac{1}{1+n}}\D_G^{\frac{1}{2}\left(\frac{n+2}{n+1}\right)}\, .
\ee  
With the help of Eq.~(\ref{eq:rhoG}) the minimum condition in Eq.~(\ref{eq:condmin}) becomes
\be
  R \ll \frac{\mathcal{K}}{c_g}e^{-\frac{1}{2}\mathcal{W}\left(\frac{1}{2}\left(\frac{2+n}{1+n}\right)\mathcal{K}^2\right)}\simeq \frac{\mathcal{K}}{c_g}\ll 1\, .\label{eq:minGinv}
\ee 
Furthermore, the last equalities hold for $\mathcal{K} \ll 1$ and $c_g\gg 1$ which will turn out to be a good approximation for typical parameter values (see \S\ref{SCcases}). 

As we will demonstrate explicitly in \S\ref{SCcases}, the minimum solution in practice has a very limited range of validity, since the evolution turns out the be non-adiabatic very close to the origin.

\subsubsection{Non-Adiabatic Solution for the Gaussian Core Profile}  
In the non-adiabatic case, in the core region, $0\leq R \leq A$, the field equation takes the form
\be
  \frac{ {\rm d}^2 \phi}{{\rm d} R^2}+\frac{2}{R}\frac{ {\rm d}\phi}{{\rm d} R}\simeq \frac{\beta}{\M}\rho_{\rm max}e^{-\frac{1}{2}(Rc_g)^2}
\ee
with
\be
  \rho_{\rm max}=\frac{\tilde{\rho}}{A^2}e^{\frac{1}{2}(Ac_g)^2}\, ,
\ee
where the last equality results from matching Eq.~(\ref{eq:rhoIS}) and Eq.~(\ref{eq:rhoG}) at $R=A$.

Requiring $\left.{\rm d}\phi/{\rm d}R\right|_{R=0}=0$, with the help of Eq.~(\ref{eq:phiSCdiffin}) and Eq.~(\ref{eq:Deltar2}) for $0\leq R \leq A$ we arrive at
\ba
  \phi_{\rm G}(R)&=&\phi_{\rm free}-\D_{\rm G}\left[\sqrt{\frac{\pi}{2}}\frac{{\rm erf}(\frac{1}{\sqrt2}Rc_g)}{Rc_g}
    -\left(1+(Ac_g)^2\log(A)\right)e^{-\frac{1}{2}(Ac_g)^2}\right],\label{eq:phiGin}\\
  \D_{\rm G}&=&\frac{\beta\rho_{\rm max}}{\M}\left(\frac{r_{\rm sc}}{c_g}\right)^2=\frac{e^{\frac{1}{2}(Ac_g)^2}}{(Ac_g)^2}\D_{\rm IS},\\
  c_g&=& \frac{1}{\sqrt{-1-\log(A)}A}\label{eq:Acg},
\ea
where the relation in Eq.~(\ref{eq:Acg}) guarantees continuity of $\phi$ and its derivative at $R=A$.

We now have an analytical solution for $\phi_{\rm G}$ which allows us to determine the slope of the isothermal profile at $R=A$.

\subsubsection{Isothermal Distributions with a Gaussian Core Region}
\label{ISGauss}
In this section, we determine the dependence of $\langle (\nabla \phi)^2 \rangle$ in the isothermal region on the slope of the field profile in the core region. For this purpose, we require the scalar field profile and its derivative to be continuous at the matching point $A$. Recall that this condition fixes the relation in Eq.~(\ref{eq:Acg}).

Taking the derivative of the field profile in Eq.~(\ref{eq:phiGin}) at $R=A$ and inserting it for $\left.{\rm d}\phi/{\rm d}R\right|_{R=A}$ in Eq.~(\ref{eq:phiSCdiffin}), we arrive at
\ba
  \phi_{\rm{in}}&=&\phi_{\rm{free}}-\D_{\rm IS}\left[1-\log(R)-\frac{A}{R}\chi\right],\label{eq:phiinSC}\\
  \phi_{\rm{out}}&=&\phi_{\rm{free}}-\D_{\rm IS}\left[1-A\chi\right]\frac{e^{-m_{\rm{free}}r_{SC}(R-1)}}{R}\, ,\nonumber\label{eq:SCphiout}
\ea
where
\be
  \chi\equiv\left(1+\frac{1-{\rm erf}\left(\frac{Ac_g}{\sqrt{2} }\right)e^{\frac{1}{2}(A c_g)^2}}{(Ac_g)^2}\right)\nonumber.
\ee
The scalar field profile in Eq.~(\ref{eq:phiinSC}) leads to a gradient squared averaged over the volume of the object of
\ba
  \langle (\nabla \phi)^2 \rangle&=&3\left(\frac{\D_{\rm IS}}{r_{\rm SC}}\right)^2\left[ 1+\left(\frac{2\log(A)}{A\chi}-1\right) A^2\chi^2+(\chi^2-1)A\right] \nonumber \\
   &\simeq&3\left(\frac{\D_{\rm IS}}{r_{\rm SC}}\right)^2\left[1+\left(\frac{4}{3}\log(A)-\frac{5}{9}\right)A+\mathcal{O}(A^2)\right],\label{eq:gradsqSC}
\ea
where in the last equality we have employed Eq.~(\ref{eq:Acg}) and assumed $A\ll 1$. Noting that for $\chi \rightarrow 1$ we recover the result in Eq.~(\ref{eq:phiSCdiffin}) with $A\left.{\rm d}\phi/{\rm d}R\right|_{R=A} \ll\D_{\rm IS}$ , we are now in a position to estimate the impact of the field derivative at the matching point on the average gradient squared. For $A\ll 1$ and $\log(A)\ll 0$, the field gradient at $R=A$ leads to an increase of the result by a factor $\sim (1-2/3\log(A)A) >0$ which amounts to ${\mathcal{O}}(15\%)$ for typical parameter values.

Accordingly, we find that a small central region with (approximately) Gaussian matter distribution has a sizeable effect on the scalar field dynamics. This has to be contrasted with the negligible impact of the corresponding NFW matter distribution at small radii which was found to result from its cusped structure.

For the volume average of the potential in the case of an exponential potential we get 
\be
  \langle V \rangle \simeq V_{\rm free}\left[1+\left(\frac{4}{3}-A\right)\frac{\alpha\D_{\rm NFW}}{\M}\right].\label{eq:VexpavSC}
\ee
In the case of an inverse power law we arrive at
\be
  \langle V \rangle \simeq V_{\rm free}\left[ 1+\left(\frac{4}{3}-A\right)n\frac{\D_{\rm NFW}}{\phi_{\rm free}}\right].\label{eq:VinvavSC}
\ee
In both cases, the quintessence potential could be expanded around the free-field potential for $\alpha\D/\M\ll 1$ and $\D/\phi_{\rm free}\ll 1$, respectively. As in the case of the NFW profile in \S\ref{NFW}, we find a negligible correction term. 

Note that our result can be generalised to any analytic quintessence potential which allows for such an expansion. So long as $\D$ is much smaller than the characteristic mass scale of the scalar potential, the correction term will remain negligible. 

In summary, for all compact objects from galactic up to supercluster scales, the absolute change in the scalar field value induced by the scalar/matter coupling has a negligible effect on the average potential of the scalar field. Taken across the Hubble volume, this implies that possible corrections to the homogeneous average equation of state can directly be traced back to the square of non-trivial field gradients. Note also that as a result of the small present value of $\overline{V}\simeq V_{\rm free}\simeq 10^{-120}\M^4$, the field gradients only have to be of comparable magnitude in order to induce sizeable effects on the average equation of state.
  
\subsection{Supercluster Case Studies: Analytical and Numerical Results}
\label{SCcases}
In this section we discuss our analytical and numerical results for SC $8$ and SC $98$, which we take as representatives for further and less evolved superclusters respectively. The difference in their evolutionary state is reflected in the shape of their spherically averaged density profiles, plotted in Fig.~\ref{Profiles} together with a Gaussian (for $R\ll 1$) and an isothermal profile (for $R\leq A$) fitted to the data. The distribution of SC $8$ drops faster than $R^{-2}$ at $R\sim 0.5$; in the case of SC $98$ the slope is already changing at a much smaller radius, $R \sim 0.1$. Note also that SC $8$ has a larger mass than SC $98$ and therefore its impact on the the scalar field value is larger. This is demonstrated in Fig.~\ref{Phis}, which shows the scalar field profile for the two superclusters according to Eq.~(\ref{eq:phiGin}) and Eqs.~(\ref{eq:phiinSC}--\ref{eq:SCphiout}).

We are now in a position to verify the limited range of validity of the adiabatic solution according to Eqs.(\ref{eq:minGexp}) and (\ref{eq:minGinv}). For our test superclusters we arrive at
\be
  \D_{\rm G}=1.8\times 10^{-2}\M
\ee
for SC $8$, and
\be
  \D_{\rm G}=5.8\times 10^{-3}\M
\ee
for SC $98$. Therefore, for both cases, the adiabaticity condition is already violated at $R \ll 1$ such that in practise the field evolution is non-adiabatic.

In the following, we will compare and discuss our analytical and numerical results for $Q$ as defined in Eq.~(\ref{eq:wavsum}) and analyse the impact of deviations from the isothermal profile at large radii. Finally, based on our results, we will estimate $\overline{w}$.

Taking Fig.~\ref{Profiles} as a rough guideline for the relevant scales, according to Eq.~(\ref{eq:phiinSC}) we find for the isothermal matter distribution that
\be
  Q=\frac{\langle (\nabla \phi)^2 \rangle}{V_{\rm free}}\simeq 1.35\beta^2\left(\frac{r_{\rm sc}}{30h^{-1}\rm{Mpc}}\right)^2\left(\frac{\tilde{\rho}}{42\rho_{c}}\right)^2.
\ee

\begin{center}
\FIGURE{
\hspace*{-15pt}
\includegraphics[width=0.54\textwidth]{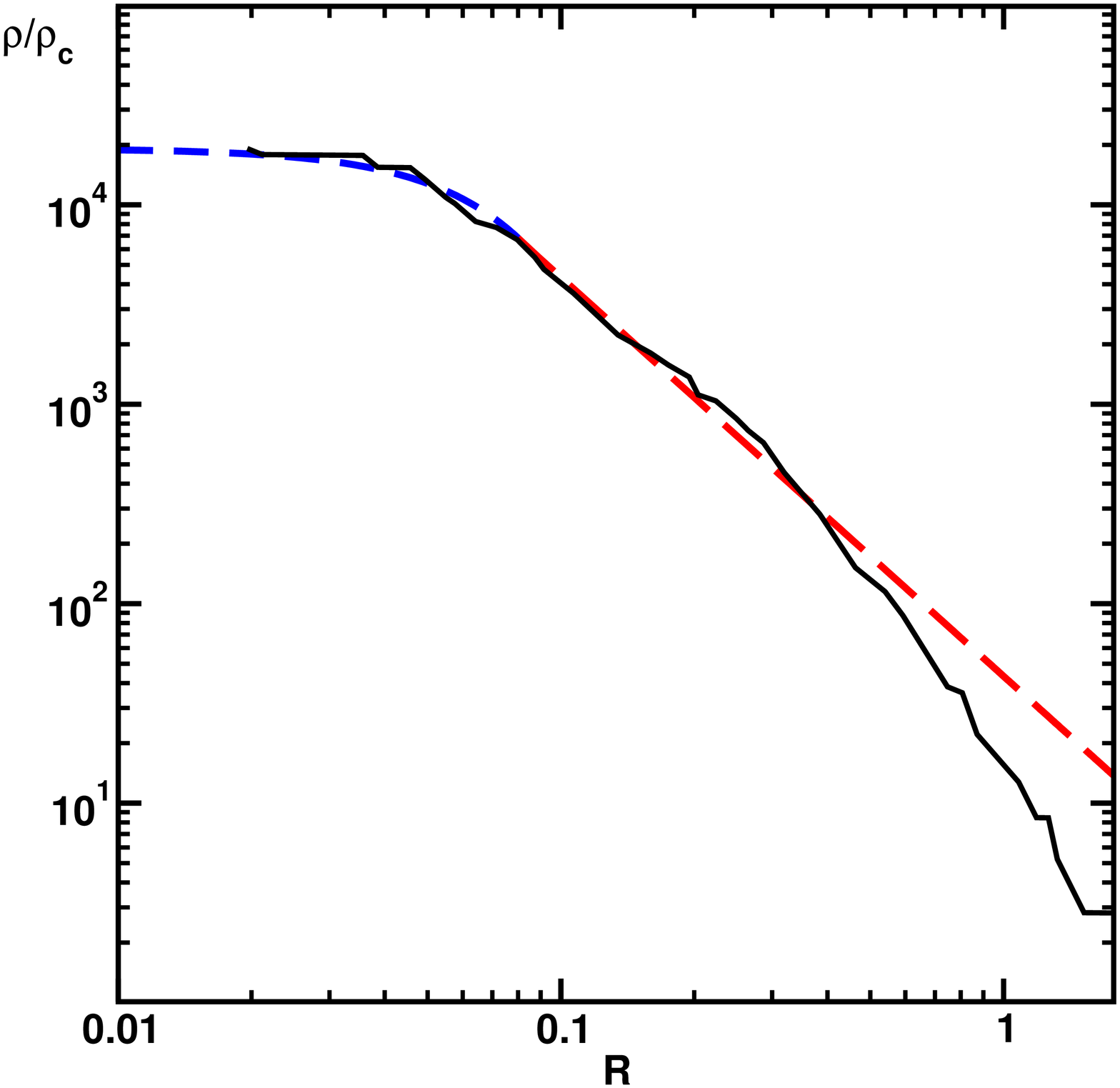}\includegraphics[width=0.54\textwidth]{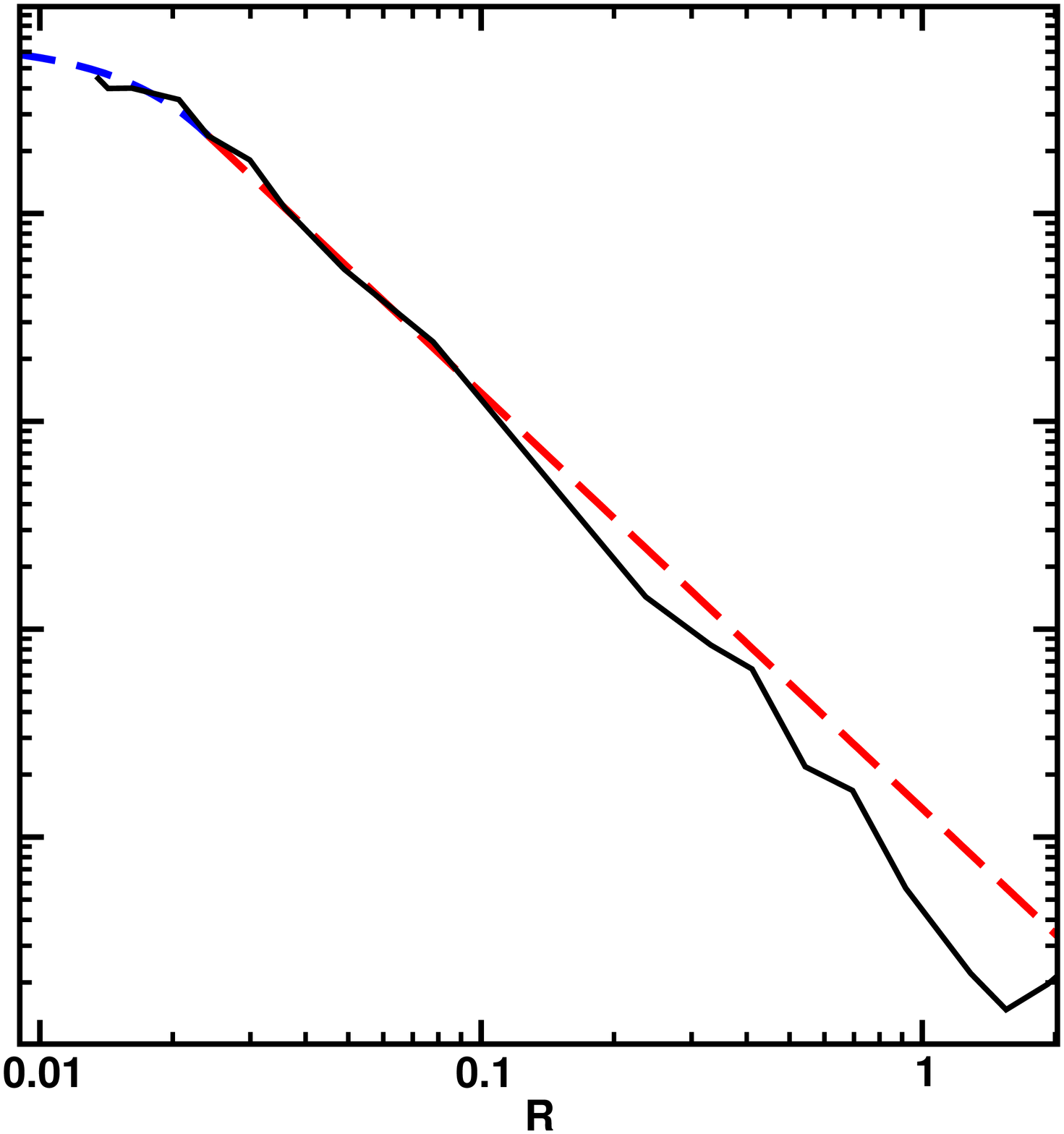}
\caption{Spherically averaged matter density profiles of the further and less far evolved superclusters $8$ and $98$, respectively. Short dashed lines denote the Gaussian core region, long dashed lines the $R^{-2}$ isothermal tail, and solid black lines the numerical results.}
\label{Profiles}
}
\end{center}

\begin{center}
\FIGURE{
\qquad\includegraphics[width=0.54\textwidth]{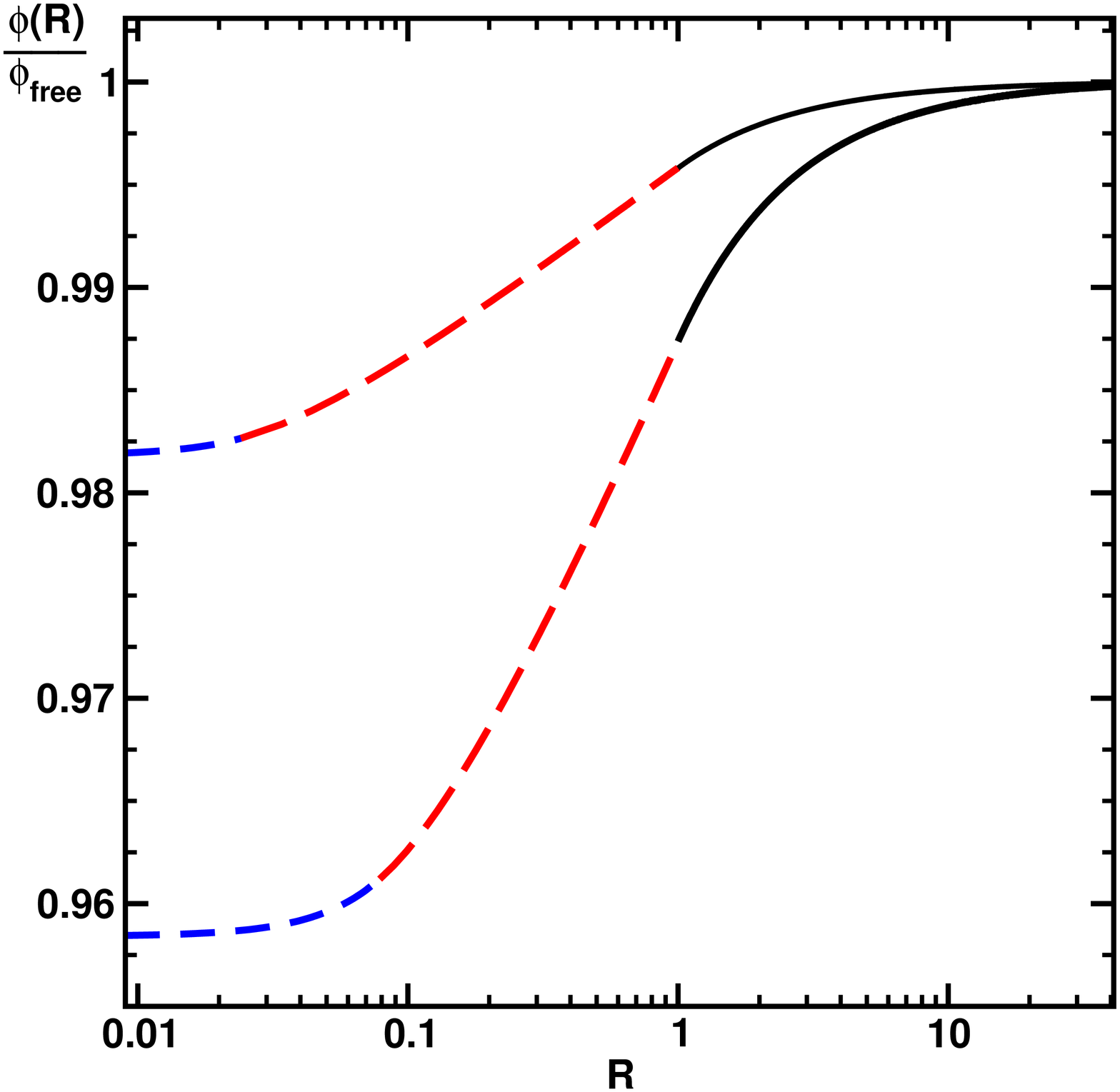}\qquad
\caption{The scalar field profile $\phi(R)/\phi_{\rm free}$ for the analytic fits for SC $8$ (lower curve) and SC $98$ (upper curve) and $\beta=1$. The solutions in the Gaussian region are plotted with short dashed lines, those in the isothermal region with long dashed lines, and the external solutions with solid black lines.}
\label{Phis}
}
\end{center}

To compare this result with our numerical findings, we insert the fitted radial density distributions shown in Fig.~\ref{Profiles} into the general solution for the average gradient squared in Eq.~(\ref{eq:gradgen})\footnote{Note that we have extended the data at small radii with the analytical data from the Gaussian fit in order to employ the initial condition in Eq.~(\ref{eq:ini1}) for $\nabla\phi$ at $R=0$}. Fig.~\ref{GavsQs} shows our numerical and analytical results for the normalised average gradient squared,
\be
  Q(R)=V_{\rm free}^{-1}\int \limits^{R}_{0}{\rm d} \tilde{R}\left(\nabla \phi\right)^2\tilde{R}^2\, ,\label{eq:QR}
\ee
and its integrand of as a function of $R$. As demonstrated in the lower panels of Fig.~\ref{GavsQs}, the final value of the normalised average gradient squared for SC $8$ is $Q_8=1.63$, almost an order of magnitude larger than that for SC $98$, where $Q_{98}=0.21$. Furthermore, for the former the analytical and numerical curves are in very good agreement for all radii, while in the latter case the analytical result overestimates $Q(R)$ even at small radii. As can be understood from the plots in the upper panel of Fig.~\ref{GavsQs} of the integrand of $Q$ in Eq.~(\ref{eq:QR}), the reason is that the isothermal profile overall provides a better fit to the profile of the further evolved SC $8$.

According to the analytical result discussed at the end of \S\ref{IS}, $Q$ receives almost equal contributions at all radii in the isothermal region, which is reflected by the small slope of the dashed curves. As could be expected from Fig.~\ref{Profiles}, the numerical curves agree with the analytic up to the radius at which the numerical profiles drop faster than $R^{-2}$. However, for larger radii, the numerical contributions to the integrand $Q(R)$ show a decaying behaviour. Therefore, to a good approximation the corresponding maximal value of $Q(R)$ fixes the total Q, since further contributions are negligible.

\begin{center}
\FIGURE{
\hspace*{-15pt}
\includegraphics[width=0.54\textwidth]{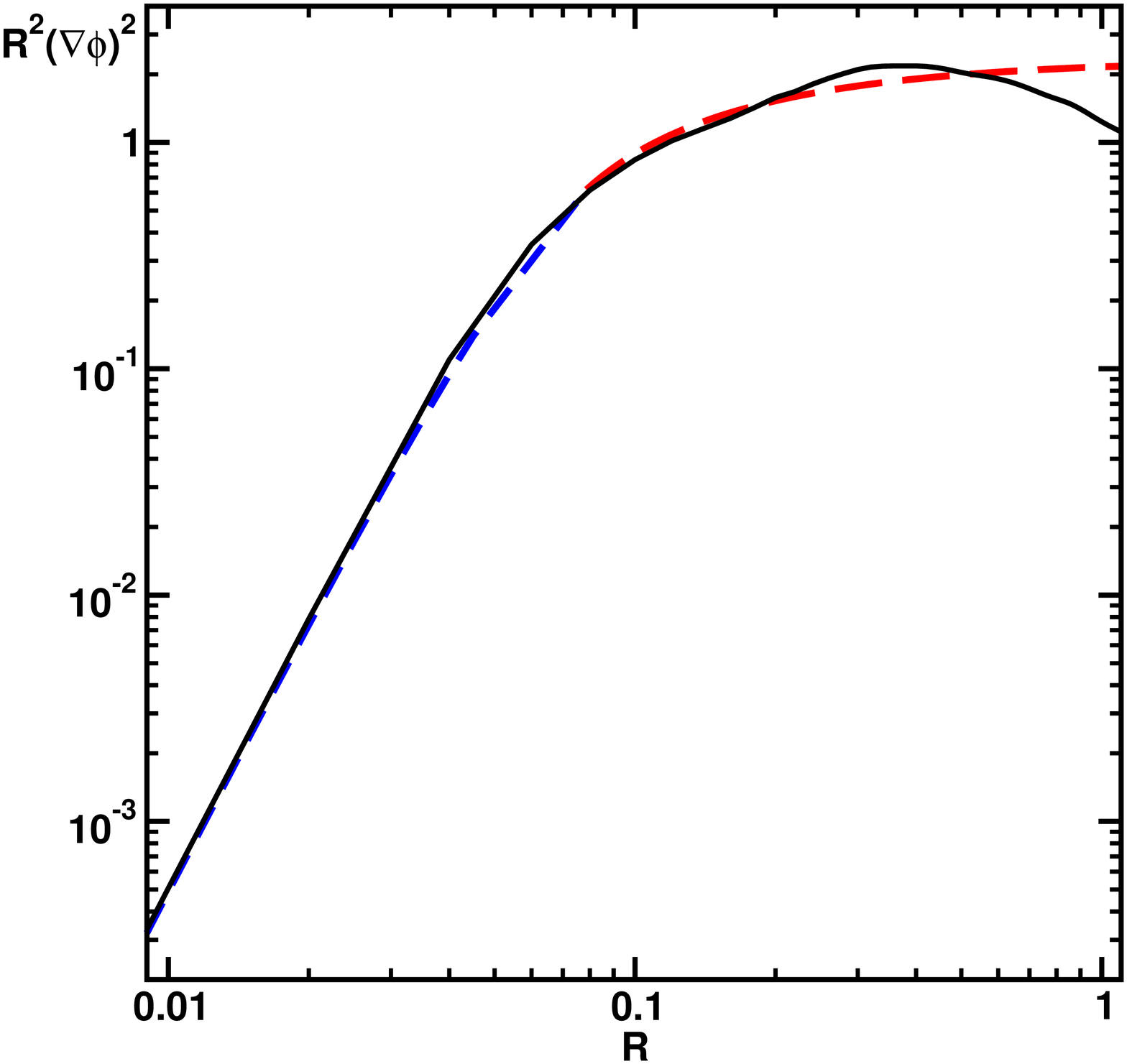}\includegraphics[width=0.54\textwidth]{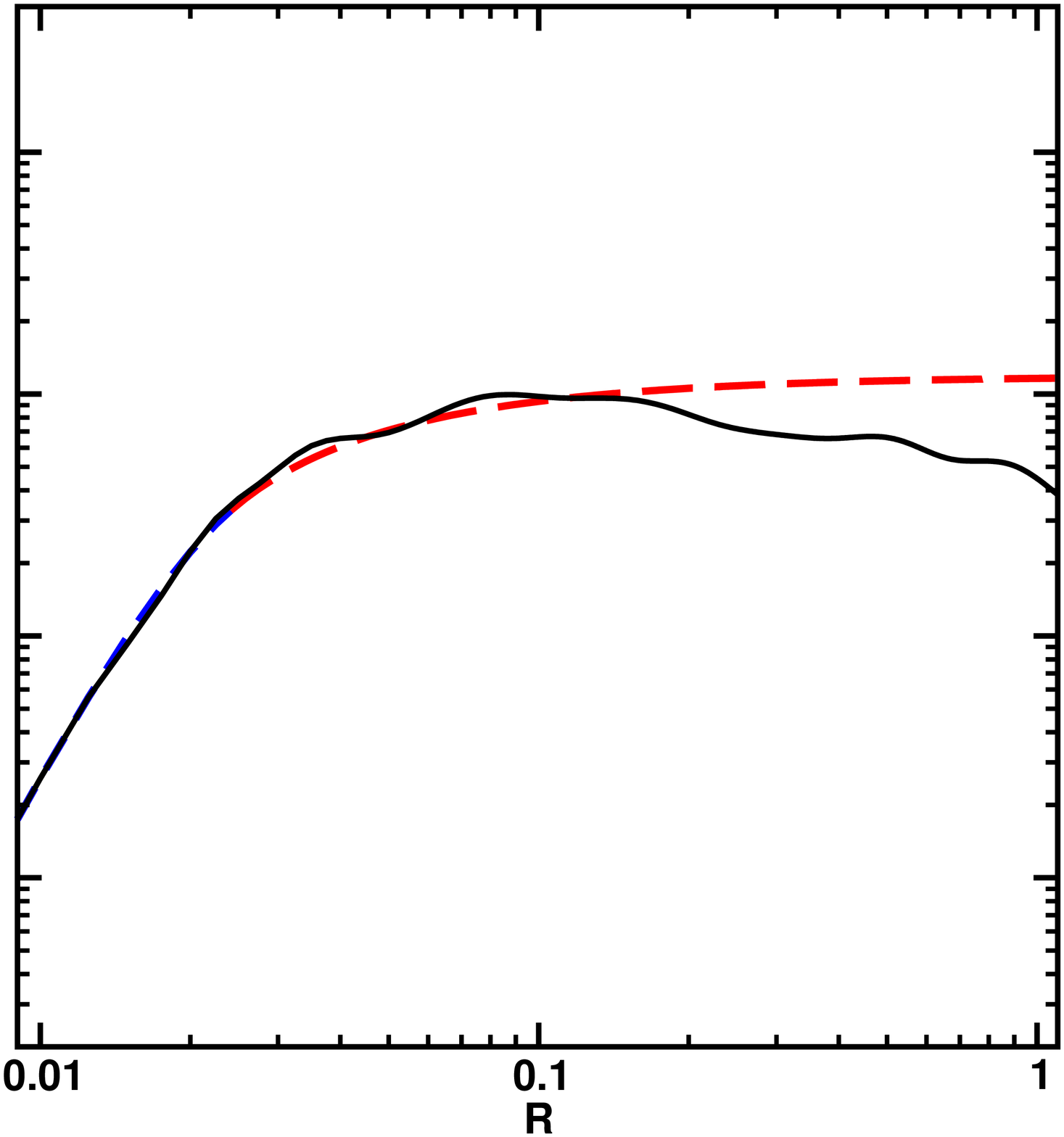}
\newline
\hspace*{-15pt}
\includegraphics[width=0.54\textwidth]{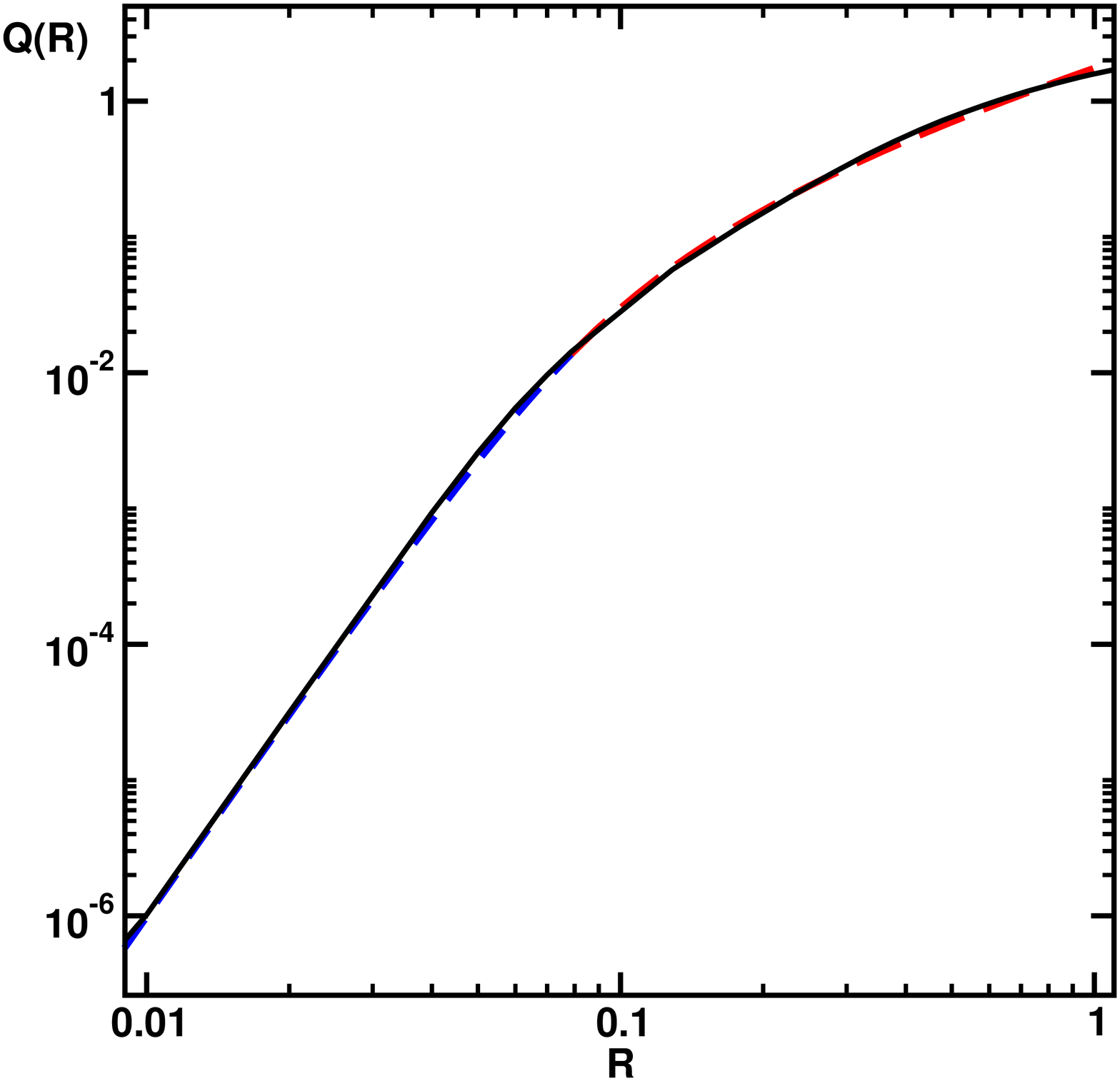}\includegraphics[width=0.54\textwidth]{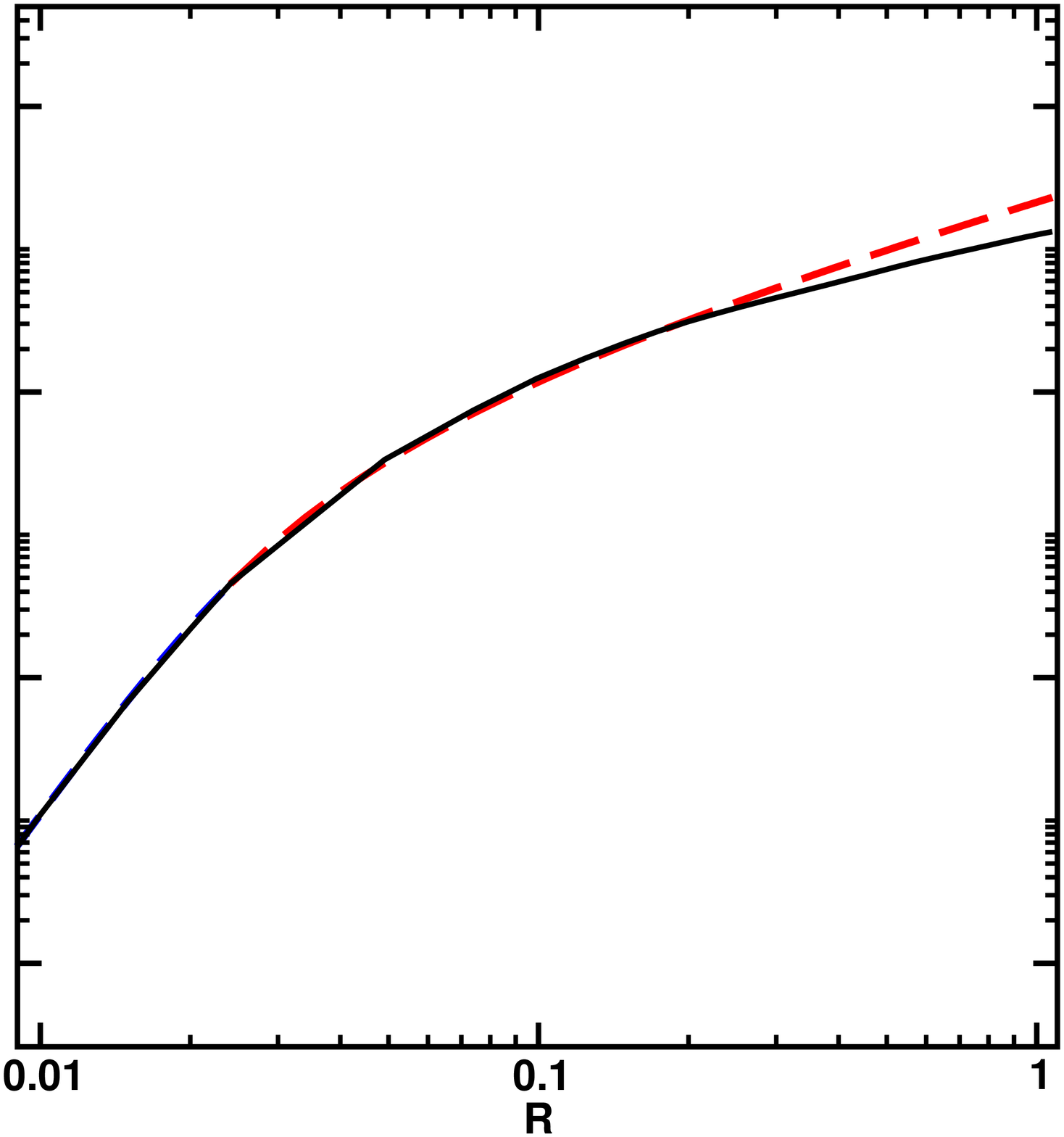}
\caption{$R^2(\nabla\phi)^2$ (top) and $Q(R)$ (bottom) for SC $8$ (left) and SC $98$ (right) and $\beta=1$.}
\label{GavsQs}
}
\end{center}

Based on our results in this section, we are finally in a position to estimate the overall effect on the expansion rate for coupled quintessence scenarios. To achieve this, we note that a particular class of high-mass superclusters presumably yields the dominant impact on the expansion rate; superclusters tend to become ever less concentrated the larger (and less evolved) they are, as reflected by their mean density,
\be
  \bar{\rho}=\frac{3}{4\pi}\frac{M_{\rm sc}}{r_{\rm sc}^3}\, ,
\ee
which implies that $\langle (\nabla \phi)^2 \rangle \propto \left(\beta/\M\right)^2M_{\rm sc}^2/r_{\rm sc}^4$. Here, the factor of proportionality depends on the radial distribution of mass within the object. Adding to this, the abundance of bound structure drops with increasing radius of the object. We therefore conclude that the main source of scalar field gradients will stem from medium-sized superclusters which are both large and numerous enough that their volume still makes up a reasonable fraction of the Hubble volume (see for comparison the discussion in \S\ref{sec:averagew} below Eq.~(\ref{eq:wavsum})).

\begin{center}
\FIGURE{
\hspace*{-15pt}
\includegraphics[width=0.54\textwidth]{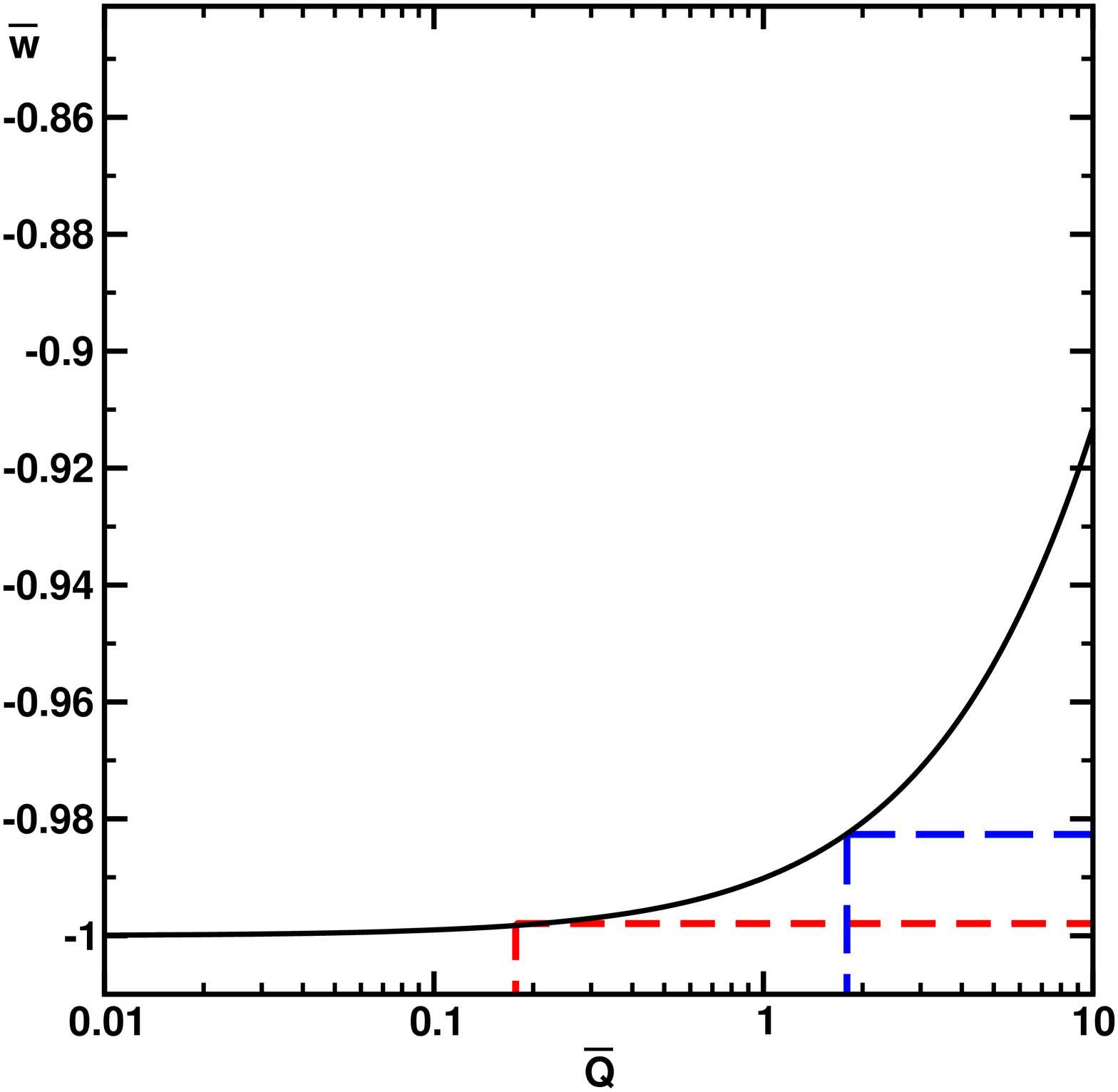}\includegraphics[width=0.54\textwidth]{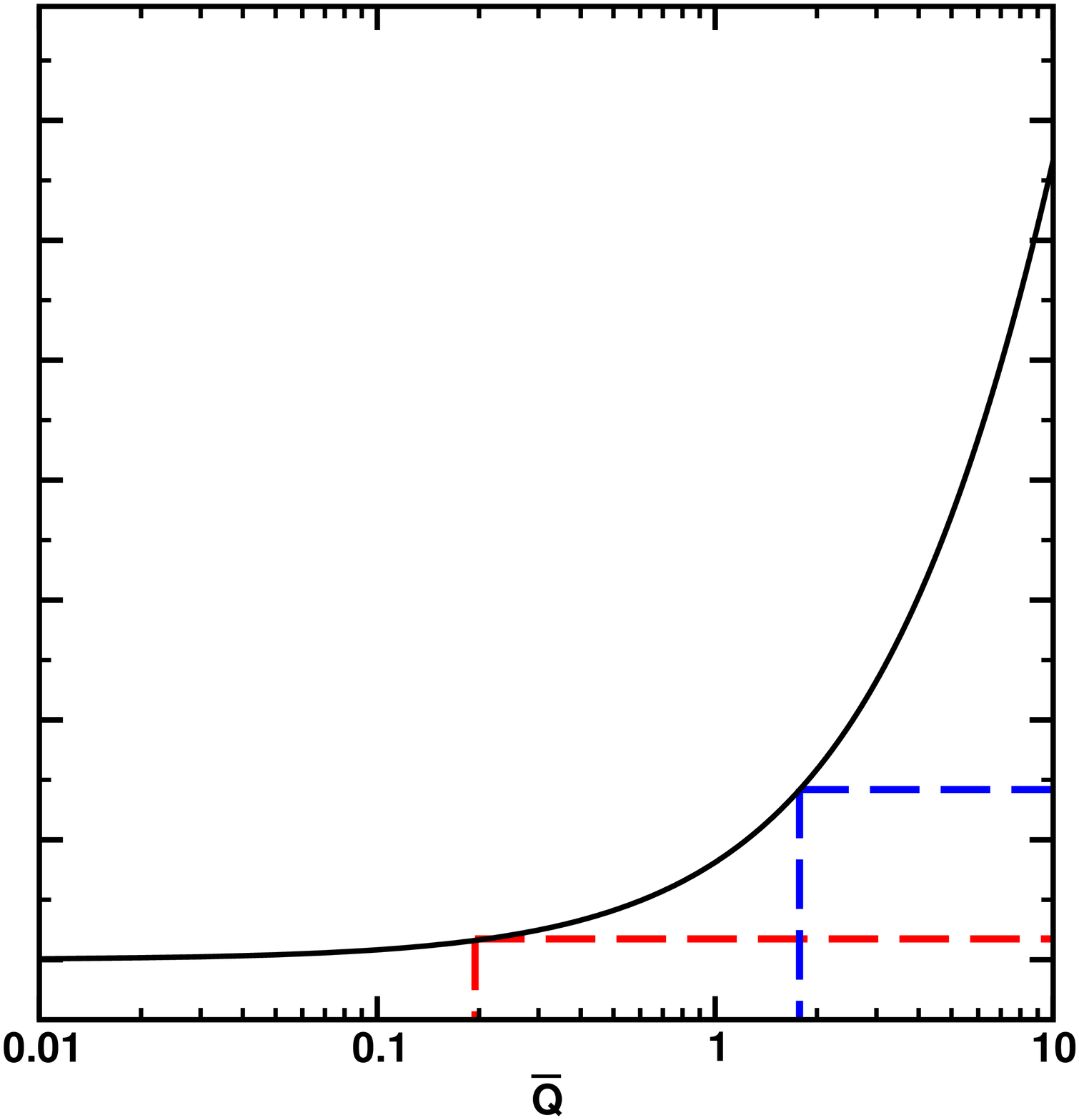}
\caption{$\overline{w}$ as a function of $\overline{Q}$ for a volume of 3\% (left) and 5\% (right) and $\beta=1$.}
\label{Ws}
}
\end{center}

As a result, we assume the total volume taken up by all superclusters ($\sim (3-5)\%\mathcal{V}_H)$ to be mainly provided by one representative class of superclusters with a medium size of $\simeq 30h^{-1}{\rm Mpc}$ and normalised gradient squared $\overline{Q}$. According to Eq.~(\ref{eq:wavsum}), this corresponds to $\mathcal{R}_i=\mathcal{R}~\sim (3-5)\times 10^{-2}$. In Fig.~\ref{Ws} we plot the resulting $\overline{w}$ for two different values of $\mathcal{R}$ as a function of $\overline{Q}$ together with the corresponding values for $\overline{w}$ which we get for superclusters similar to SC $8$ and SC $98$. 

From these plots we can see that, if superclusters similar to SC $98$ dominate, only small deviations from $\overline{w}\simeq -1$ are to be expected, even should the superclusters fill $5\%$ of the Hubble volume. However, for further evolved superclusters of type SC $8$, the deviations from the homogeneous equation of state are dramatic. For a lower abundance, where the superclusters fill $3\%$ of the universe, the equation of state is pushed to $\overline{w}\simeq -0.98$, slowing the cosmic acceleration to some degree. For the higher abundance of $5\%$, the equation of state is driven to $\overline{w}\simeq -0.97$. Let us remark that the authors of \cite{Wray:2006tw} have performed a complete analysis of a dark matter simulation taking into account the shape, multiplicity, size, and radial structure of superclusters. They found that across the entire sample the spherically averaged density profile is roughly isothermal in the range $0.3\lesssim R \lesssim 0.7$. This investigation suggests that on average, superclusters tend to be much further evolved than SC $98$ and, in fact, exhibit a similar evolutionary state to SC $8$. In light of these results, we therefore conclude that an overall correction of $\gtrsim 3\%$ to the homogenous equation of state is conceivable in scenarios with scalar/matter couplings of gravitational strength. In the next section, we comment on the class of coupled scalar field scenarios which allow for $\beta$ of $\mathcal{O}(1)$. Furthermore, in \S\ref{Outlook} we provide an outlook of the implications of our results for scenarios with couplings of super-gravitational strength such as coupled neutrino quintessence~\cite{Fardon:2003eh,Peccei:2004sz,Fardon:2005wc,Amendola:2007yx,Wetterich:2007kr}. 
In summary, as anticipated, in coupled quintessence scenarios the main source for notable corrections to $\overline{w}$ stems from the largest and most massive bound objects known. Since superclusters are typically less than two orders of magnitude smaller than the Hubble radius, the volume suppression is many orders of magnitude lower than that for galaxies and galaxy clusters.

In addition, the spatial changes in the coupled scalar field value were found to get more pronounced for higher masses of bound objects and are also sensitive to its distribution. Namely, for superclusters, the dominant isothermal region exhibits a shallower slope than the NFW profile discussed in \S\ref{NFW} and, importantly, lacks a very highly-concentrated central core (the cusp). Thus, according to their evolutionary state, the cumulative mass of an supercluster and its impact on the scalar field value change gradually over a much larger range in radius. This has to be contrasted with the fast drop in density at small radii for collapsed and virialised structures which results in a strongly suppressed contribution to the volume integral of the gradient squared. 

Therefore, we find it to be likely that in coupled quintessence scenarios the clustered matter distribution on supercluster scales has a relatively large impact on the average equation of state of dark energy, provided the interaction is of at least gravitational strength. 

It is also important to note that the results of our investigation can be taken to be conservative for the following reasons. As a simplifying assumption, we have neglected the anisotropic nature of structure formation by assuming a spherically symmetric matter density profile for the superclusters. However, the shape of structures still in the process of formation is triaxial and only becomes closer to spherical in the process of collapse and relaxation. Adding to this, while structure is infalling, significant internal substructure is still recognisable. Accordingly, within the virialised constituents of the supercluster, the scalar field value can be expected to decrease, while outside it relaxes again. Within the binding radius of the supercluster, therefore, the coupled scalar field will exhibit an oscillatory behaviour. Since the average scalar field equation of state is, importantly, only sensitive to the square of the gradient, both positive and negative scalar field gradients will act in the same way, to increase the average equation of state. Further study in this direction is underway \cite{GLI}.

\subsection{Comment on the Strength of the Scalar/Matter Coupling}
In accordance with previous numerical studies in the literature~\cite{Li:2009sy,Zhao:2009ke,Li:2010mq}, we have demonstrated analytically that in the present universe the scalar field evolution is non-adiabatic within compact structures ranging from galactic up to supercluster scales. We therefore found, for coupled quintessence models in general, that the relative importance of the average gradient squared of the scalar field compared to the average potential grows quadratically with the coupling strength $\beta$. While this result is largely independent of the scalar field potential $V(\phi)$, possible effects on the average equation of state of the quintessence field thus strongly depend on the coupling strength. In this section we comment on the implications for various coupled dark energy scenarios of the model-dependent constraints on $\beta$ that result from current cosmological observations. 

These bounds strongly depend on the dynamical evolution of the field, controlled by the interplay between the scalar potential -- and, in particular, its steepness -- and the form and strength of the scalar/matter coupling governing the effective potential $\Veff$. In the absence of a global minimum of $\Veff$, the ultra-light mass of the scalar field $m_{\phi}$ is independent of the local matter density, leading to a long-range fifth force between the coupled matter species on all scales. 

In this case, a coupling between cold dark matter and the scalar field is restricted to $5\%-20\%$ of gravitational strength by current observations\cite{Kesden:2006zb,Kesden:2006vz,Bean:2008ac}. Therefore, our results demonstrate that, even on supercluster scales, the average gradient squared of the scalar field is at least two orders of magnitude lower than the average scalar field potential. Within these coupled scenarios, deviations from the homogeneous equation of state are then negligible.
   
However, in the case when $\Veff$ allows for a minimum, by virtue of the chameleon mechanism it is possible that in high density regions the field acquires a large mass $m_{\phi}$ such that the range of the fifth force $m^{-1}_{\phi}$ becomes undetectably short~\cite{Khoury:2003aq,Khoury:2003rn,Brax:2004qh}. Stringent constraints from local tests of the equivalence principle as well as the solar system can thereby be evaded even for couplings $\beta \gg 1$~\cite{Mota:2006ed,Mota:2006fz,Brax:2007hi}. Even so, on cosmological scales it is possible that the scalar field has a low enough mass to affect the non-linear formation of matter structures in a complex manner~\cite{Li:2009sy,Zhao:2009ke,Li:2010mq}. More precisely, the scalar field acts as a potential for the fifth force, which itself depends on the underlying matter density in a different (non-linear) manner to the gravitational potential\footnote{A simple rescaling of the gravitational constant $G$ has been shown in certain regimes not to be physical~\cite{Li:2009sy,Zhao:2009ke,Li:2010mq}.}. Taking into account the spatial distribution of the scalar field for such a scenario, numerical N-body simulations have been performed to determine the impact of a scalar/matter coupling $\beta$ of gravitational strength on the power spectrum~\cite{Li:2009sy,Zhao:2009ke,Li:2010mq}, the mass function~\cite{Zhao:2009ke,Li:2010mq} and the internal profiles of dark matter halos~\cite{Li:2010mq}. One of the main conclusions is that on linear and highly non-linear scales, these strongly-coupled chameleon-like models are virtually indistinguishable from $\Lambda$CDM~\cite{Zhao:2009ke}. However, as a result of the enhanced growth in structure due to the fifth force, the matter power spectrum was found to be significantly increased with respect to $\Lambda$CDM on intermediate scales relevant for galaxy clusters. The authors of Ref.~\cite{Zhao:2009ke} conclude that future surveys operating on scales $k=0.1$--$10\,h\,{\rm Mpc}^{-1}$ will be needed if we are to further tighten constraints on the coupling below gravitational strength.

To summarise, in rather weakly coupled dark energy scenarios with $\beta\ll 1$, the effect of the scalar matter coupling on the average equation of state of the scalar field is negligible, as could be expected. However, this is not the case in chameleon-like cosmologies which allow for couplings of gravitational strength between the scalar field and dark matter or all kinds of matter. As we have demonstrated in the last section, deviations on the level of a few percent from the homogeneous equation of state are conceivable.

In the next section, we will provide an outlook for the implications of our results for even less constrained theories such as coupled neutrino quintessence~\cite{Fardon:2003eh,Peccei:2004sz,Fardon:2005wc,Amendola:2007yx,Wetterich:2007kr} with scalar neutrino couplings $\beta \gg 1$.

\section{Outlook -- Couplings of Super-Gravitational Strength}
\label{Outlook}
In this section we briefly consider the implications of our results for dynamical dark energy scenarios in which a quintessence field couples to some dark species with a coupling strength much stronger than gravity, $\beta\gg 1$, such as in coupled neutrino quintessence~\cite{Fardon:2003eh,Peccei:2004sz,Fardon:2005wc,Amendola:2007yx,Wetterich:2007kr}.

In stark contrast to a $\Lambda$CDM cosmology, the formation of structure in the coupled species does not cease in the era of dark energy domination. This is due to the fact that gravity, and therefore the time scale for gravitational collapse, is directly connected with the evolution of the scale factor. However, the attractive force mediated by the scalar field between the coupled matter is independent of the background cosmology. Consequently, it allows for structures forming up to the present time. As a direct consequence, the characteristic size of these lumps can in principle be much larger than supercluster scales, $80 h^{-1}{\rm Mpc}\ll r_{\rm obj}\lesssim 0.1{\rm H^{-1}_0}$ (depending on the specific model assumptions)~\cite{Wintergerst:2009fh}. Accordingly, even only a few of these objects can take up a very large fraction of the total Hubble volume $\mathcal{V}_H$.

\begin{center}
\FIGURE{
\hspace*{-15pt}
\includegraphics[width=0.54\textwidth]{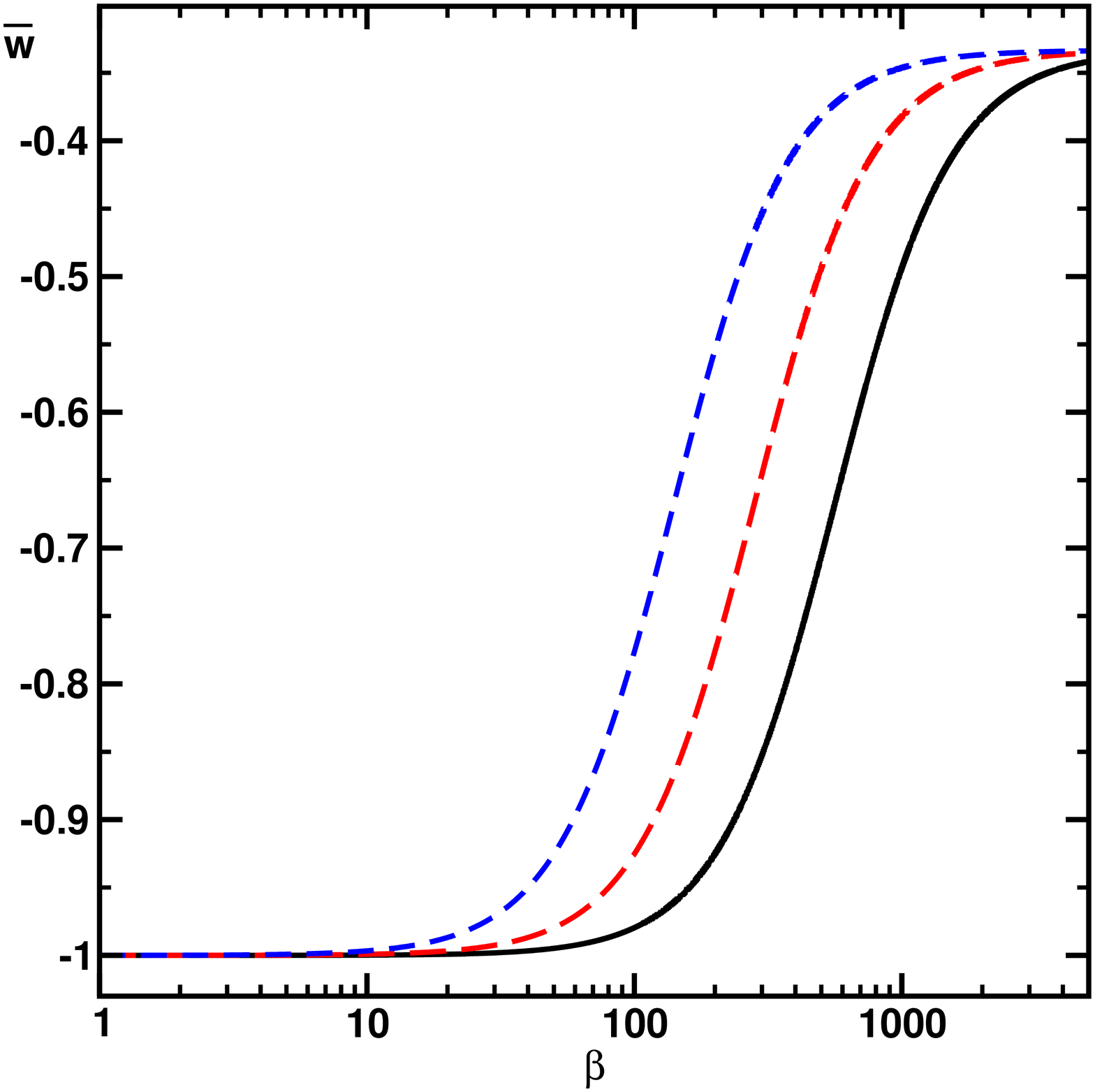}\includegraphics[width=0.54\textwidth]{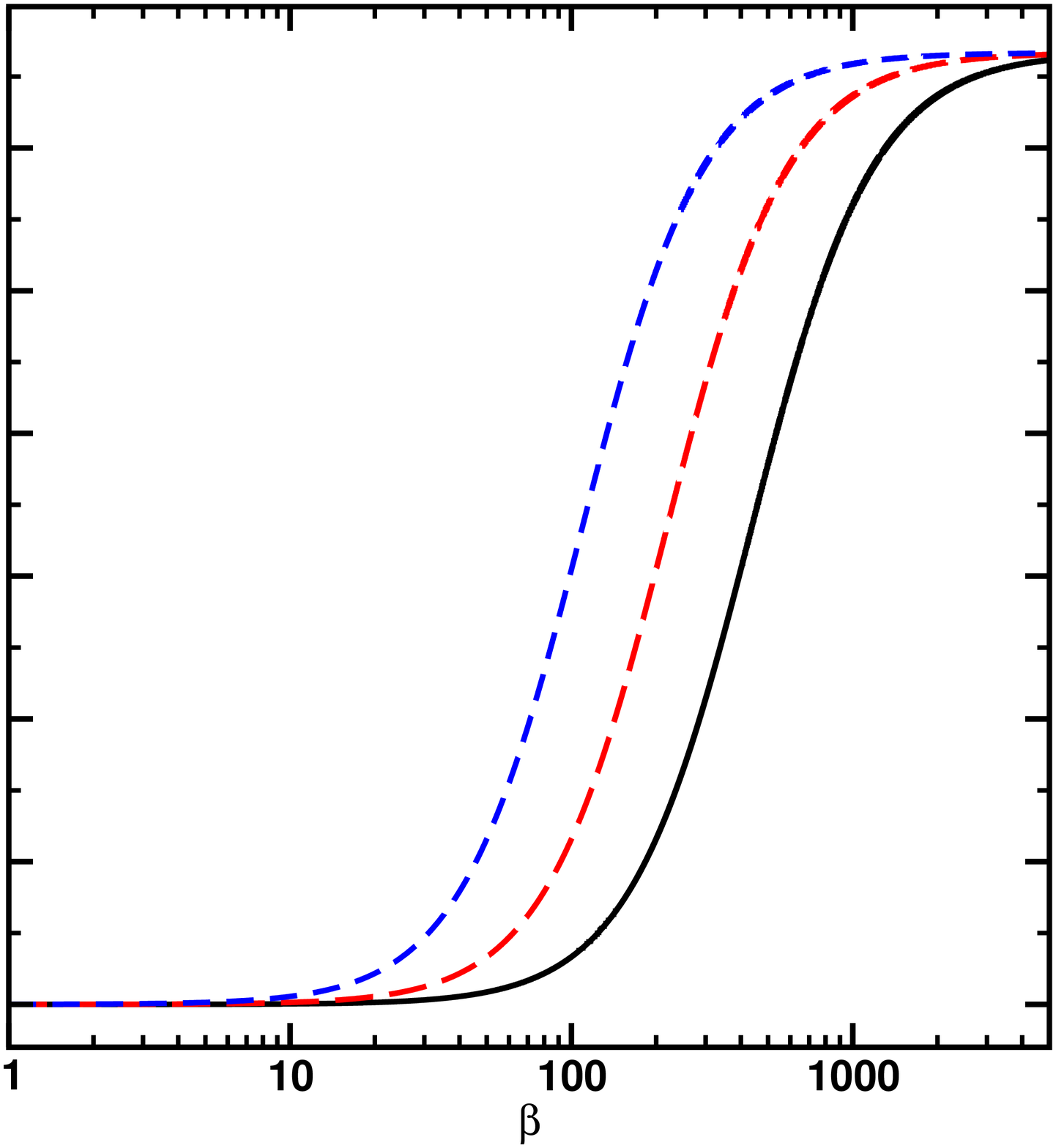}
\caption{$\overline{w}$ as a function of $\beta$ for a volume of 3\% (left) and 5\% (right) taken by bound structures with radius $r_{\rm vir}=20 h^{-1}$ Mpc and a mass of $M=5\times 10^{15}M_{\odot}$ (solid), $M=10^{16}M_{\odot}$ (long-dashed) and $M=2\times 10^{16}M_{\odot}$ (short-dashed).}
\label{WBeta}
}
\end{center}

Clearly, as in our current study, the specific functional form of the spherically averaged density profile will have a large impact on the size of possible corrections to the average equation of state of the coupled scalar field. The authors of \cite{Wintergerst:2009fh} found approximately a high-concentration NFW profile for neutrino lumps forming in a model with a high coupling strength, $|\beta|=52$ and current neutrino masses of $2$ eV. 

For highly concentrated structures, the normalised gradient squared $Q$ resulting from Eq.~(\ref{eq:gradNFW}) without employing the relations between $c$ and $r_s$ in Eq.~(\ref{eq:crs}) is
\be
  Q=\frac{\langle\left(\nabla\phi\right)^2\rangle}{V_{\rm free}}\simeq 2.28\left(\frac{\beta}{52}\right)^2\left(\frac{M_{\rm vir}}{10^{16}M_\odot}\right)^2\left(\frac{r_{\rm vir}}{20h^{-1}\rm{Mpc}}\right)^{-4}.
\ee
To get a crude estimate for the impact of such structures on the cosmic equation of state, we plot in Fig.~\ref{WBeta} the resulting $\overline{w}$ as a function of $\beta \gg 1$ for typical class of high mass objects with an NFW like matter profile. 

Even assuming an occupancy of only $3\%$, we can see that a strength $\beta\simeq 50$ will produce an average equation of state $-0.99\lesssim \overline{w}\lesssim -0.92$, depending on the ratio $M_{\rm vir}/r^2_{\rm vir}$; for an occupancy of $5\%$ the average equation of state would be $-0.99\lesssim \overline{w}\lesssim -0.88$. Therefore, compared to the case of couplings of (sub-)gravitational strength, much larger deviations to the homogeneous equation of state seem feasible in these kinds of scenarios.

In the case of an isothermal matter distribution the result can be expected to be yet larger, but it will again strongly depend on the model-dependent mass and radius of the object, as well as on its concentration. In this case it should also be verified whether the increase in $\beta$ allows for an adiabatic field evolution. 

The possibility of finding notable deviations from the homogeneous case clearly suggest that this issue warrants further study. In particular, this might imply a source of new observational constraints on the coupling strength.

\section{Conclusions}
\label{Conclusions}
In this paper, we have investigated the impact of a coupling $\beta$ between a quintessence field and clustered matter on the average scalar field equation of state across the universe $\overline{w}$. For this purpose, we have derived its general form which turns out to crucially depend on the ratio of the average gradient squared to the average potential of the field. In the static limit, this ratio vanishes for a perfectly homogeneous uncoupled scalar field, implying $\overline{w}\simeq -1$. However, this is not the case in the presence of sizeable spatial variations in the field, where in the static limit the equation of state could reach an asymptotic value of $\overline{w}\simeq -1/3$.

As a first step, we considered isolated bound objects with non-trivial density profiles characteristic of their size and evolutionary state. Solving the equations of motion for the field analytically for a variety of field potentials, we determined the spatial profile of a quintessence field induced by the coupling to matter. To model objects on galactic and cluster scales, we assumed a matter profile of the NFW form, while on supercluster scales we considered an isothermal matter distribution $\propto r^{-2}$ modified at small and large radii.

A first crucial result is that, within bound objects on all scales, the absolute change in the value of the coupled scalar field is much smaller than the fundamental field value which is of order $\M$. An immediate consequence is that, for a typical quintessence potential, the average of the scalar field potential over the volume of \emph{any} bound object is largely independent of the strength of the coupling. Furthermore, this implies that the correction to the uncoupled value of the potential is negligible.

However, the average gradient squared of the field grows quadratically with the strength of the coupling. Importantly, its magnitude furthermore exhibits a strong dependence on the functional form of the matter distribution; in the case of the NFW profile, the corresponding volume integral is strongly suppressed both at small and at large radii. As a result, the average gradient squared is at least a factor $\mathcal{O}(10^2)\beta^2$ smaller than the scalar field potential, even for the largest and most massive galaxy clusters.

More significant results can be found in the case of galaxy superclusters. Fully evolved superclusters obey an isothermal density profile across much of their radii, proportional to $r^{-2}$. The average gradient squared consequently receives almost equal contributions at all radii and, as a result, for typical parameters it can be at least $\beta^2$ times larger than the scalar field potential.

This result depends crucially on the shape of the profile; however, unlike fully virialised objects such as galaxies and galaxy clusters, superclusters are still forming. Their spherically averaged density profile then depends on their evolutionary state and will deviate from the isothermal at small and large radii. To probe the dependence of the results on the evolutionary state of the superclusters we considered two representative test objects from simulations, one relatively unevolved and one much more so.

In both cases, we found the matter in the core region to be well-fitted by a Gaussian distribution, while at large radii the slope of the distribution dropped faster than $r^{-2}$. We built an analytical model for the density profile, stitching a Gaussian core to an extended isothermal tail. We could then solve the equations of motion both analytically and numerically, allowing the following general conclusions. At small and medium radii, the contributions to the average gradient squared of the coupled field resulting from the analytical fit agree very well with the findings from the numerical data. However, as soon as the matter profile drops faster than $r^{-2}$, the integrand of the volume integral is no longer approximately constant, but instead decays. Accordingly, further evolved superclusters result in larger average field gradients, since the isothermal profile provides a better fit for a larger range in radius; in our particular cases the average gradients squared were approximately an order of magnitude greater for the more evolved object than for the less evolved. Adding to this, for a power-law density profile, the magnitude of the average field gradient squared grows quadratically with the supercluster mass, but is suppressed by the fourth power of the radius.

Taken across the Hubble volume, these results suggest that the greatest corrections to the homogeneous equation of state in coupled quintessence scenarios are expected to originate from further evolved, medium sized superclusters; these objects are both large and numerous enough to fill a reasonable fraction of the Hubble volume while additionally inducing relatively large deviations from the homogeneous behaviour. We could therefore infer from our supercluster case studies that a scalar/matter coupling of gravitational strength allows for an increase in the average equation of state of the universe of up to $\sim 3\%$. Let us emphasise that this is a conservative estimate, since substructures will presumably induce oscillations in the field gradients; both positive and negative gradients will increase the result.

In coupled dark energy scenarios which allow for couplings of gravitational strength, the global equation of state of the scalar field across the Hubble volume will therefore be displaced from $\overline{w}\approx -1$. As it is this equation of state that one would expect to employ in the Friedmann equations, the global acceleration rate in coupled quintessence models will therefore be less than would be na\"ively expected.

It should also be noted that particular coupled quintessence scenarios which allow for super-gravitational strength couplings -- to neutrinos, for example -- can produce even larger deviations, since the average gradient squared grows quadratically with the coupling strength. We briefly considered this scenario using the NFW profile and found even for this distribution that corrections across the entire Hubble volume of up to $\sim 10\%$ seem feasible for typical model parameters.

\acknowledgments{The authors wish to thank Javier Grande for important contributions in the early stages of this work and an anonymous referee for helpful suggestions.}

\bibliographystyle{JHEP}
\bibliography{Bib}

\end{document}